\documentclass[11pt,preprint]{aastex}
\usepackage{lscape}

\begin{document}

\title{Chemical composition of the old globular clusters
NGC~1786, NGC~2210 and NGC~2257 in the Large Magellanic Cloud.
\footnote{Based on observations obtained 
at Paranal ESO Observatory under proposal  080.D-0368(A)}}

\author{Alessio Mucciarelli}
\affil{Dipartimento di Astronomia, Universit\`a 
degli Studi di Bologna, Via Ranzani, 1 - 40127
Bologna, ITALY}
\email{alessio.mucciarelli2@unibo.it}

\author{Livia Origlia}
\affil{INAF - Osservatorio Astronomico di Bologna, Via Ranzani, 1 - 40127
Bologna, ITALY}
\email{livia.origlia@oabo.inaf.it}

\author{Francesco R. Ferraro}
\affil{Dipartimento di Astronomia, Universit\`a 
degli Studi di Bologna, Via Ranzani, 1 - 40127
Bologna, ITALY}
\email{francesco.ferraro3@unibo.it}

\begin{abstract}

This paper presents the chemical abundance analysis of a sample of 18 giant 
stars in 3 old globular clusters in the Large Magellanic Cloud, namely 
NGC~1786, NGC~2210 and NGC~2257. The derived iron content is 
[Fe/H]=~--1.75$\pm$0.01 dex 
($\sigma$=~0.02 dex), ~--1.65$\pm$0.02 dex 
($\sigma$=~0.04 dex) and ~--1.95$\pm$0.02 dex 
($\sigma$=~0.04 dex) for NGC~1786, NGC~2210 and NGC~2257, respectively. 
All the clusters exhibit similar abundance ratios, with 
enhanced values ($\sim$+0.30 dex) of [$\alpha$/Fe], consistent with 
the Galactic Halo stars, thus indicating that these clusters have formed from 
a gas enriched by Type II SNe. 
We also found evidence that {\sl r}-process are the main channel
of production of the measured neutron capture elements (Y, Ba, La, Nd, Ce and Eu). 
In particular the quite large enhancement of [Eu/Fe] ($\sim$+0.70 dex) found in these 
old clusters clearly indicates a relevant efficiency of the {\sl r}-process 
mechanism in the LMC environment.
\end{abstract}

\section{Introduction}

In the last decade, the advent of the high resolution spectrographs 
mounted on the 8-10 m telescopes
has allowed to extend the study of the chemical composition of 
individual Red Giant Branch (RGB) stars outside our Galaxy 
up to dwarf and irregular galaxies of the Local Group.
Chemical analysis of RGB stars are now available for 
several isolated dwarf spheroidal (dSph) galaxies 
as Sculptor, Fornax, Carina, Leo I, Draco, Sextans and Ursa Minor 
\citep{shetrone01, shetrone03, letarte} and 
the Sagittarius (Sgr) remnant \citep{boni00, monaco05, monaco07, sbordone}. 
As general clue, these studies reveal that the chemical abundance patterns 
in the extragalactic systems do not resemble those observed in the Galaxy, 
with relevant differences 
in the [$\alpha$/Fe]
\footnote{We adopt the usual spectroscopic notations that 
[$X_1$/$X_2$]=~$\lg(N_{X_{1}}/N_{X_{2}})_{*}$ - $\lg(N_{X_{1}}/N_{X_{2}})_{\odot}$
and that $\lg(N_{X_{1}})$=~$\lg(N_{X_{1}}/N_{H})$+12.},
[Ba/Fe] and [Ba/Y] ratios, thus suggesting 
different star formation history and chemical evolution
\citep[see e.g.][]{venn, geisler, tolstoy}.

Unlike the dSphs, the irregular galaxies as the Large Magellanic 
Cloud (LMC) contain large amount of gas and dust, showing an efficient 
ongoing star-formation activity. The LMC globular clusters (GCs)
span a wide age/metallicity range, with both old, metal-poor 
and young, metal-rich objects, due to its quite complex star formation 
history. Several events of star formation occurred: the first one 
$\sim$13 Gyr ago and 4 main bursts at later epochs, 
2 Gyr, 500 Myr, 100 Myr and 12 Myr ago \citep{harris09}.
Until the advent of the new generation of spectrographs, the study 
of the chemical composition 
of the LMC stars was restricted to red and blue supergiants 
\citep{hill95, korn00, korn02}, providing information only about 
the present-day 
chemical composition. The first studies based on high resolution 
spectra of RGB stars \citep{hill00, johnson06, pompeia08} provided 
first and crucial  information about the early chemical enrichment and 
nucleosynthesis.

Of the $\sim$300 compact stellar clusters listed by \citet{kon}, 
metallicity determinations from Ca~II triplet are available 
for some tens of objects \citep{ols91,aaron} and only for 7 clusters 
high-resolution spectroscopic analysis have been carried out 
\citep{hill00, johnson06}. 
With the final aim of reconstructing the formation history 
of star clusters in the LMC, a few years ago we started a systematic 
spectroscopic screening of giants in a sample 
of LMC GCs with different ages.

In the first two papers of the series \citep{f06,m08} 
we presented the chemical analysis of 20 elements 
for 4 intermediate-age LMC clusters 
(namely, NGC~1651, 1783, 1978, 2173).
Moreover, \citet{m09} 
discussed the iron content and the abundances of
O, Na, Mg and Al for 3 old LMC clusters 
(namely NGC~1786, 2210 and 2257), discovering 
anticorrelation patterns similar to those observed in Galactic 
clusters.
Here we extend the abundance analysis to additional 13 chemical 
elements in these 3 LMC clusters, also performing a detailed 
comparison with stellar populations in our Galaxy (both in the 
field and in globulars) and in nearby dSphs. 

The paper is organized as follows: Section~2 presents the dataset 
and summarized the adopted procedure to derive radial 
velocities; Section~3  describes the methodology used to infer 
the chemical abundances; Section~4 discusses the uncertainties 
associated to the chemical abundances. Finally, Section~5 and ~6 
present and discuss the results of the chemical analysis.

\section{Observational data}

\begin{figure}[h]
\plotone{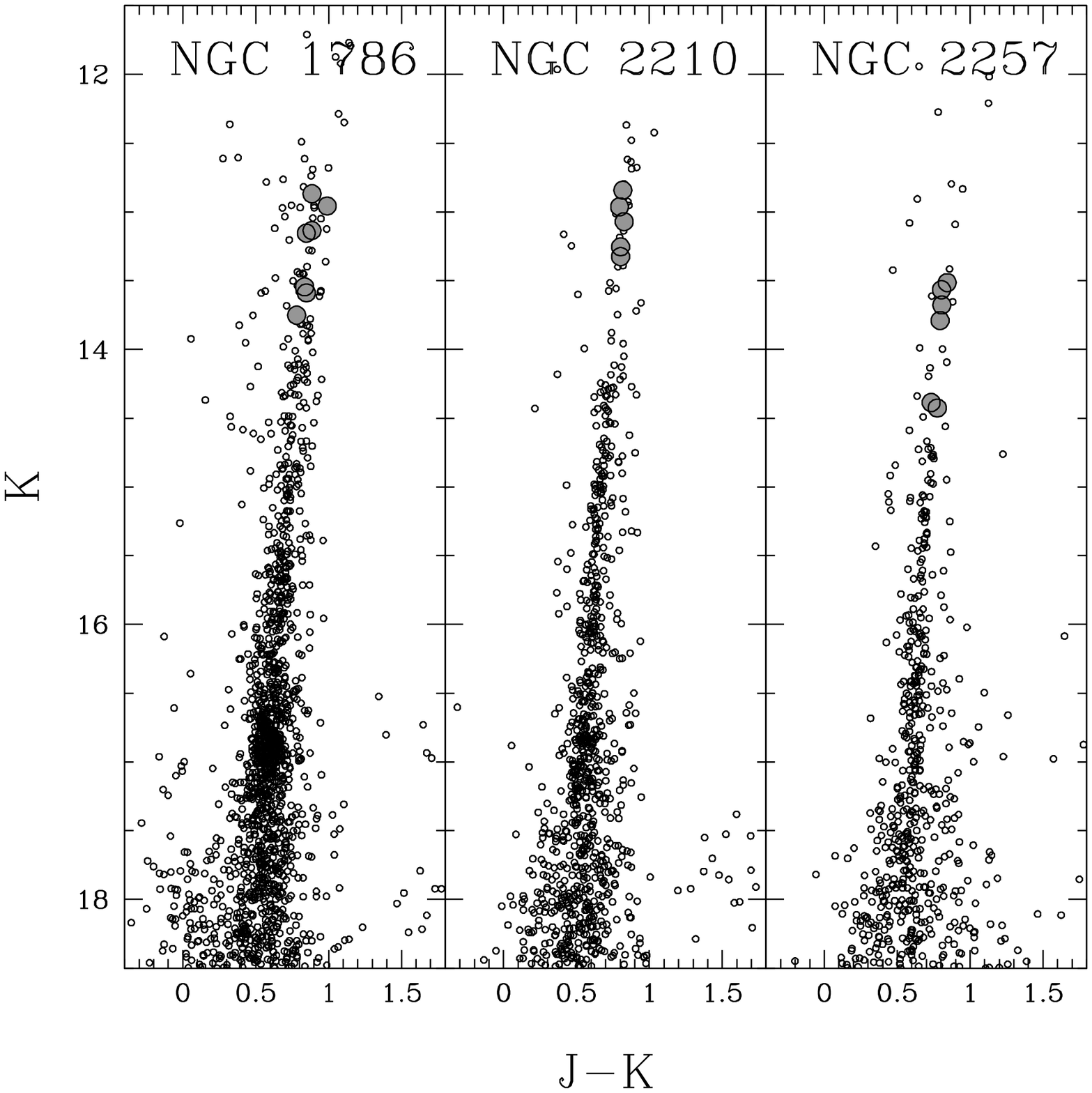}
\caption{Color-Magnitude Diagrams in the (K, J-K) plane 
of the 3 LMC old clusters: grey points indicate the stars observed with 
FLAMES.}
\label{cmd}
\end{figure}

The observations were carried out 
with the multi-object spectrograph FLAMES \citep{pasquini02} 
at the UT2/Kuyeen ESO-VLT (25-27 December 2007). 
We used FLAMES in the UVES+GIRAFFE combined mode, 
feeding 8 fibers to the UVES high-resolution spectrograph
and 132 to the GIRAFFE mid-resolution spectrograph. 
The UVES spectra have a wavelength coverage between 
4800 and 6800 $\mathring{A}$ with a spectral 
resolution of $\lambda/\Delta\lambda\sim$45000. 
We used the following GIRAFFE gratings: HR~11 
(with a wavelength range between 5597 and 5840 
$\mathring{A}$ and a resolution of $\sim$24000) 
and HR~13 (with a wavelength range between 6120 and 6405 
$\mathring{A}$ and a resolution of $\sim$22000). 
These 2 setups have been chosen in order to measure 
several tens of iron lines, $\alpha$-elements and to sample 
Na and O absorption lines. 
Target stars have been selected on the basis of (K, J-K) 
Color-Magnitude Diagrams (CMDs), as shown in Fig.~\ref{cmd},
from near infrared observations performed with SOFI@NTT 
(A. Mucciarelli et al. 2010, in preparation).
For each exposure 2 UVES and a ten of GIRAFFE fibers 
have been used to sample the sky and allow an accurate 
subtraction of the sky level.

The spectra have been acquired in series of 8-9 
exposures of $\sim$45 min each and pre-reduced independently 
by using the UVES and GIRAFFE ESO pipeline 
\footnote{http://www.eso.org/sci/data-processing/software/pipelines/}, 
including bias subtraction, flat-fielding, 
wavelength calibration, pixel re-sampling and spectrum extraction. 
For each exposure, the sky spectra have been combined together;
each individual sky spectrum has been checked to exclude possible 
contaminations from close stars. Individual stellar spectra have been 
sky subtracted by using the corresponding median sky spectra, then 
coadded and normalized. We note that the sky level is only a few percents 
of the stars level, due to brightness of our targets, introducing 
only a negligible amount of noise in the stellar spectra. 
Note that the fibre to fibre relative transmission has been taken 
into account during the pre-reduction procedure.
The accuracy of the wavelength calibration has been checked 
by measuring the position of some telluric OH and $O_2$ emission lines 
selected from the catalog of \citet{oster}.

\subsection{Radial velocities}
Radial velocities have been measured by using the IRAF 
\footnote{Image Reduction and Analysis facility. IRAF is 
distributed by the National Optical Astronomy Observatories, 
which is operated by the association of Universities for 
Research in Astronomy, Inc., under contract with the 
National Science Foundation.}
task FXCOR, performing a cross-correlation between the observed spectra 
and high S/N - high resolution spectrum of a template star of 
similar spectral type. For our sample we selected a K giant 
(namely HD-202320) whose spectrum is available 
in the ESO UVES Paranal Observatory Project database
\footnote{http://www.sc.eso.org/santiago/uvespop/} \citep{bagnulo}. 
Then, heliocentric corrections have been computed 
with the IRAF task RVCORRECT. 
Despite the large number of availables fibers, only a few
observed stars turned out to be cluster-member, due to the small 
size of the clusters within the FLAMES field of view. 
We selected the cluster-member stars according to their radial velocity, 
distance from the cluster center and position on the CMD.
Finally, we identified a total of 7 stars in NGC~1786, 5 stars in 
NGC~2210 and 6 stars in NGC~2257.
We derived average radial velocities of 
$V_{r}$=~264.3 km $s^{-1}$ ($\sigma$=~5.7 km $s^{-1}$), 
337.5 km $s^{-1}$ ($\sigma$=~1.9 km $s^{-1}$) and 
299.4 km $s^{-1}$ ($\sigma$=~1.5 km $s^{-1}$) for 
NGC~1786, 2210 and 2257, respectively. The formal error 
associated to the cross-correlation procedure is 
of $\sim$0.5--1.0 km $s^{-1}$.
The derived radial velocities are consistent 
with the previous measures, both from integrated spectra 
\citep{dubath} and from low/high-resolution individual stellar spectra 
\citep{ols91, hill00, aaron}.
In fact, for NGC~1786 \citet{ols91} estimated 
264.4 km $s^{-1}$ ($\sigma$=4.1  km $s^{-1}$) from 
2 giant stars, while \citet{dubath} 
provide a value of 262.6 km $s^{-1}$.
For NGC~2210 the radial velocity 
provided by \citet{ols91} is of 342.6 km $s^{-1}$ ($\sigma$=7.8  km $s^{-1}$), 
while \citet{dubath} and \citet{hill00} obtained radial velocities
of 338.6 and 341.7 km $s^{-1}$ ($\sigma$=2.7  km $s^{-1}$), respectively.
For NGC~2257, \citet{aaron} provided a mean value of 301.6
km $s^{-1}$ ($\sigma$=3.3  km $s^{-1}$)
and \citet{ols91} of 313.7 km $s^{-1}$ ($\sigma$=2.1  km $s^{-1}$).
For all the targets Table~\ref{info} lists the S/N computed 
at 6000 $\mathring{A}$ for the UVES spectra and 
at 5720 and 6260 $\mathring{A}$ for the GIRAFFE-HR11 and 
-HR13 spectra, respectively. Also, we report 
$V_{r}$, dereddened $K_0$ magnitudes and $(J-K)_0$ colors 
and the RA and Dec coordinates (onto 2MASS 
astrometric system) of each targets.

\section{Chemical analysis}
Similarly to what we did in previous works \citep{f06,m08,m09}, 
the chemical analysis has been carried out  
using the ROSA package (developed by R. G. Gratton, 
private communication). 
We derived chemical abundances 
from measured equivalent widths (EW) of single, unblended lines, 
or by performing a $\chi^2$ minimization between observed and synthetic 
line profiles for those elements (O, Ba, Eu) for which 
this approach is mandatory (in particular, to take into account the 
close blending between O and Ni at 6300.3 $\mathring{A}$ and the 
hyperfine splitting for Ba and Eu lines).
We used the solar-scaled Kurucz model atmospheres with overshooting 
and assumed that local thermodynamical equilibrium (LTE) 
holds for all species. Despite the majority of the available abundance analysis 
codes works under the assumption of LTE, 
transitions of some elements are known to suffer from 
large NLTE effects \citep{asplund}. 
Our Na abundances were corrected for these effects 
by interpolating the correction grid computed by \citet{gratton99}.

The line list employed here 
is described in details in \citet{gratton03} and \citet{gratton07} 
including transitions for which accurate laboratory and 
theoretical oscillator strengths are available and has been 
updated for some elements affected by hyperfine structure and isotopic splitting.
Eu abundance has been 
derived by the spectral synthesis of the Eu~II line at 
6645 $\mathring{A}$, in order to take into 
account its quite complex hyperfine structure, with a 
splitting in 30 sublevels. Its hyperfine components 
have been computed using the code LINESTRUC, described by 
\citet{wal05} and adopting the
hyperfine constants A and B by \citet{law1} and the meteoritic 
isotopic ratio, being Eu in the Sun built predominantly 
through r-process. For sake of homogeneity we adopted 
the log~gf by \citet{bkm} already used in \citet{m08}
instead of the oscillator strength by \citet{law1}.
Ba~II lines have relevant hyperfine structure components 
concerning the odd-number isotopes $^{135}$Ba and $^{137}$Ba, while the 
even-number isotopes have no hyperfine splitting; moreover, there are
isotopic wavelength shifts between all the 5 Ba isotopes. In order to 
include these effects, we employed the linelist for the Ba~II lines 
computed by \citet{prochaska} that adopted a r-process isotopic mixture. 
We note that the assumption of the r-process isotopic mixture instead of 
the solar-like isotopic mixture is not critical for the 3 Ba~II lines 
analyzed here (namely, 5853, 6141 and 6496 $\mathring{A}$), because such an effect 
is relevant for the Ba~II resonance lines 
\citep[see Table 4 by][]{sneden96}.

For the La abundances we have not taken into account 
the hyperfine structure because the observed lines are 
too weak (typically 15-30 m$\mathring{A}$) and located in the 
linear part of the curve of growth where the hyperfine 
splitting is negligible, changing the line profile but 
preserving the EW.  
Abundances of V and Sc include corrections for
hyperfine structure obtained adopting the linelist 
by \citet{whal} and \citet{pro}.\\ 
In a few stars only upper limits 
for certain species (i.e. O, Al, La and Ce)
can be measured. For O, upper limits have been 
obtained by using synthetic spectra \citep[as described in][]{m09}, 
while for Al, La and Ce computing the 
abundance corresponding to the minimum measurable EW (this latter 
has been obtained as 3 times the uncertainty derived by the 
classical Cayrel formula, see Section \ref{eq}).

As reference solar abundances, we adopted the ones computed by 
\citet{gratton03} for light Z-odd, $\alpha$ and iron-peak elements, 
using the same linelist employed here. For neutron-capture 
elements \citep[not included in the solar analysis by][]{gratton03} we
used the photospheric solar values by \citet{gs98}. 
All the adopted solar values are 
reported in  Tables \ref{el1}, \ref{el2} and \ref{el3}.

\subsection{Equivalent Widths}
\label{eq}
All EWs have been measured by using
an interactive procedure developed at our institute.
Such a routine allows to extract a spectral region 
of about 15-25 $\mathring{A}$ around any line of interest.
Over this portion of spectrum we apply a $\sigma$-rejection 
algorithm to remove spectral lines and cosmic rays. 
The local continuum level for any line 
has been estimated by the peak of the 
flux distribution obtained over the surviving points 
after the $\sigma$-rejection. 
Finally the lines have been fitted with a gaussian profile 
(rejecting those lines with a FWHM strongly discrepant with respect 
to the nominal spectral resolution or with flux residuals asymmetric 
or too large) and 
the best fits are then integrated over the selected region to 
give the EW.
We excluded from the analysis lines with $\lg{(EW/\lambda)}<$--4.5, 
because such strong features can be dominated by the contribution
of the wings and too sensitive to the velocity fields.
We have also rejected
lines weaker than $\lg{(EW/\lambda)}$=--5.8 because they are too noisy.

In order to estimate the reliability and uncertainties 
of the EW measurements, we performed some sanity checks by using 
the EWs of all the measured lines, excluding only
O, Na, Mg, and Al lines, due to their intrinsic 
star-to-star scatter (see \citet{m09} and Sect.\ref{result}):

\begin{itemize}
\item The classical formula by \citet{cayrel88} provides an 
approximate method to estimate the 
uncertainty of  EW measurements, as a function 
of spectral parameters (pixel scale, FWHM and S/N). 
For the UVES spectra, we estimated an uncertainty of  
1.7 m$\mathring{A}$ 
at S/N=~50 , while for the GIRAFFE 
spectra an uncertainty of 2 m$\mathring{A}$ at S/N=~100. 
As pointed out by \citet{cayrel88} this estimate should be 
considered as a lower limit for the actual EW uncertainty, 
since the effect of the continuum determination is not included.

\item In each cluster we selected a pair of stars with similar 
atmospheric parameters and compared the EW measured
for a number of absorption lines in the UVES spectra. The final 
scatter (obtained diving the dispersion by $\sqrt{2}$) turns 
out to be 5.6, 8.3 and 7.6 m$\mathring{A}$ for NGC~1786, 2210 and 2257, 
respectively. 

\item We compared the EWs of two target stars with similar 
atmospherical parameters observed with UVES (NGC~1786-1248) and 
GIRAFFE (NGC~1786-978), in order to check possible 
systematic errors in the EW measurements due to the use of 
different spectrograph configurations. We found 
a scatter of 6.5 m$\mathring{A}$. Within the uncertainties 
arising from the different S/N conditions and the small numbers statistic, 
we do not found relevant systematic discrepancies between the EWs 
derived from the two different spectral configurations.

\end{itemize}
\subsection{Atmospherical parameters}
 
Table~\ref{tabparam} lists the adopted atmospherical parameters 
for each target stars and the corresponding [Fe/H] abundance ratio. 
The best-model atmosphere for each target star
has been chosen in order to satisfy simultaneously 
the following constraints:\\ 
(1)~ $T_{eff}$ must be able to well-reproduce the 
excitation equilibrium, without any significant trend between 
abundances derived from neutral iron lines and 
the excitation potential;\\ 
(2)~ log~g is chosen by forcing the difference between 
log~N(Fe~I) and log~N(Fe~II) to be equal to the solar value, 
within the quoted uncertainties;\\ 
(3)~ the microturbulent velocity ($v_t$) has been obtained 
by erasing any trend of Fe I lines abundances with their expected 
line strengths, according with the prescription of \citet{magain84};\\ 
(4)~ the global metallicity of the model must reproduce 
the iron content [Fe/H];\\ 
(5)~ the abundance from the Fe~I lines should be constant with 
wavelength.

Initial guess values $T_{eff}$ and log~g have been computed 
from infrared photometry, obtained with SOFI@NTT 
(A. Mucciarelli et al. 2010, in preparation).
Effective temperatures were derived from dereddened $(J-K)_0$ colors by means 
of the $(J-K)_0$-$T_{eff}$ calibration by \citet{alonso99,alonso01}. 
The transformations between photometric systems have been 
obtained from \citet{carpenter} and \citet{alonso99}.
For all the target clusters we adopted the reddening values reported by 
\citet{persson}. Photometric gravities have been calculated from the 
classical equation:
$$log~(g/g_{\odot}) = 4\cdot log(T_{eff} / T_{eff,\odot}) + log~(M/M_{\odot}) - 0.4\cdot (M_{bol}-M_{bol,\odot})$$
by adopting the solar reference values according to IAU recommendations \citep{anders}, 
the photometric $T_{eff}$ , a distance modulus of 18.5 and a mass value of M=0.80 $M_{\odot}$, 
obtained with the isochrones of the Pisa Evolutionary Library \citep{cariulo} for 
an age of 13 Gyr and a metal fraction of Z=~0.0006.

The photometric estimates of the atmospherical parameters have been optimized 
spectroscopically following the procedure described above.
Generally, we find a good agreement between the photometric and spectroscopic 
$T_{eff}$ scales, with an average difference $T_{eff}^{spec}$-$T_{eff}^{phot}$=~-14 K 
($\sigma$=~59 K) and only small adjustments were needed 
(for sake of completeness we report in Table~\ref{tabparam} both the 
spectroscopic and photometric $T_{eff}$).
Changes in gravities are of $\pm$0.2-0.3 dex, consistent within the uncertainty 
of the adopted stellar mass, distance modulus and bolometric corrections. 

\begin{figure}[h]
\plotone{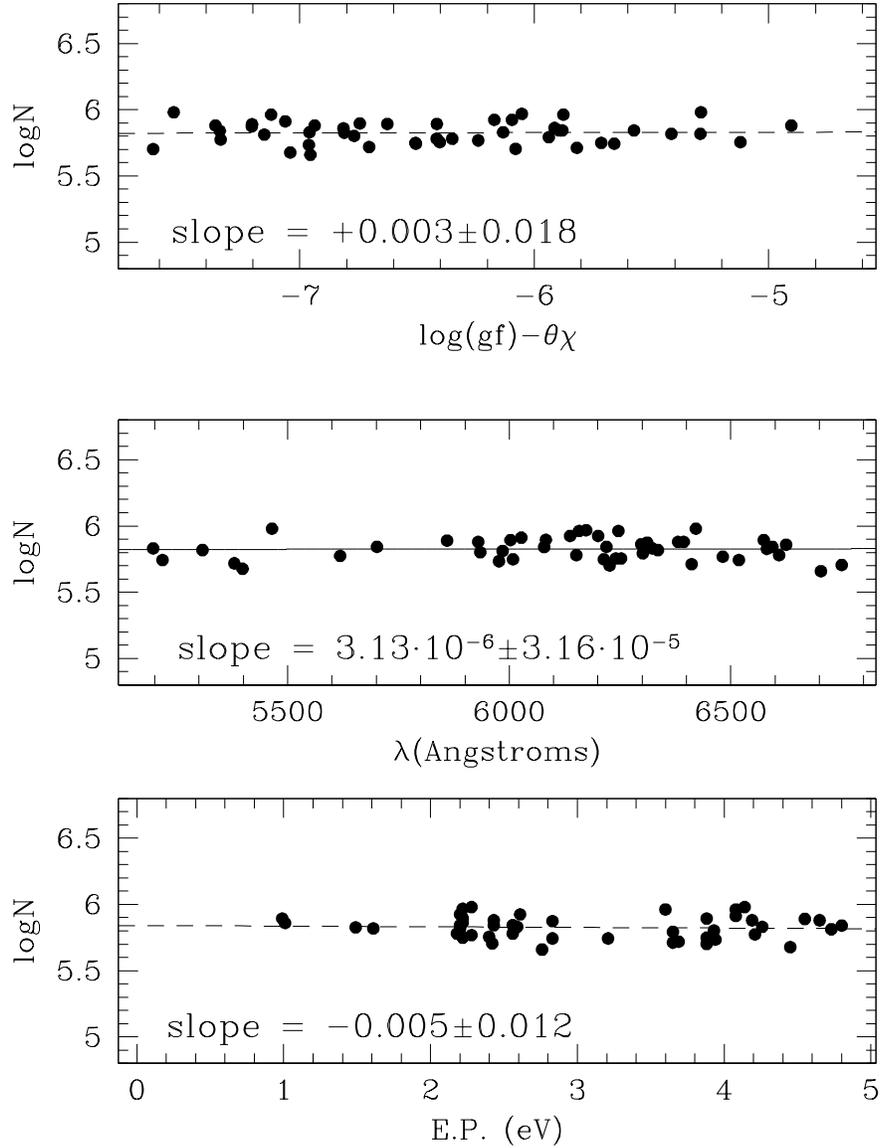}
\caption{The behavior of the number density abundance of the neutral 
iron lines as a function of the expected line strength (upper panel), 
of the wavelength (middle panel) and of the excitational potential 
(lower panel). In each panel is also reported the linear best-fit (dashed lines)
and the corresponding slope (with associated error) is labelled.}
\label{trend}
\end{figure}

An example of the lack of spurious trends between 
the Fe~I number density and the expected line strength, the 
wavelength and the excitational potential is reported in 
Fig.~\ref{trend} (linear best-fits and the corresponding slopes 
with associated uncertainties are labeled).

\section{Error budget}
\label{err}

In the computation of errors, we have taken into account 
the random component related mainly to the EW measurement uncertainty and 
the systematic component due to the atmospheric parameters. 
The total uncertainty has been derived as the sum in quadrature of random 
and systematic uncertainties. 

{\sl (i)~Random errors.}~
Under the assumption that each line provides an independent indication 
of the abundance of a species, the line-to-line scatter normalized to the root 
mean square of the observed lines number ($\sigma/\sqrt{N_{lines}}$) 
is a good estimate of the random error, arising mainly from the uncertainties 
in the EWs (but including also secondary sources of uncertainty, as the 
line-to-line errors in the employed log~gf). 
Only for elements with less than 5  available lines, we 
adopted as random error the line-to-line scatter obtained from 
the iron lines normalized for the root mean square of the number of 
lines. These internal errors are reported in Tables \ref{tabparam} - \ref{el3} 
for each abundance ratio and they are of the order of 0.01--0.03 dex for 
[Fe/H] (based on the highest number of lines) and range from $\sim$0.02 dex 
to $\sim$0.10 dex for the other elements. 

{\sl (ii)~Systematic errors.}~
The classical approach to derive the uncertainty due to the choice of 
the atmospherical parameters is to re-compute the abundances by altering each parameter 
of the corresponding error and fixing the other quantity each time. 
Then, the resulting abundance differences are summed in quadrature, providing 
the total uncertainty. 
In the case of our analysis, where the spectroscopic method to infer the 
parameters has been adopted, $T_{eff}$, log~g and $v_t$ turn out to be not 
independent each other. Variations of $T_{eff}$ affect in different ways Fe I and 
Fe II abundances, and imply related changes in log~g to compensate. 
Moreover, strongest lines have typically lower excitation potential, and any change 
in $T_{eff}$ requires a change in $v_t$. 
The method to sum in quadrature the abundance uncertainties 
under the assumption that $T_{eff}$, log~g and $v_t$ are uncorrelated is 
unable to take into account the covariance terms 
due to the dependencies among the atmospherical parameters. 
The risk to use this technique, when the spectroscopical optimization is adopted, 
is to overestimate this source of error, providing only a conservative upper limit, 
especially in cases of abundances with relevant covariance terms.\\ 
A more realistic estimate of the effective error due to the atmospherical 
parameters, can be obtained with the procedure described by \citet{cayrel04}. 
We repeated the analysis of a target star (namely, NGC~1786-2310, chosen as 
representative of the entire sample) varying $T_{eff}$ by $\pm$100 K 
with respect to the best model $T_{eff}$ and repeating the entire 
procedure to optimize the other parameters, deriving new best values 
for log~g and $v_t$: we obtained log~g=~0.9 and $v_t$=~2 km $s^{-1}$ 
when we increase $T_{eff}$ of 100 K, and log~g=~0.3 and $v_t$=~1.85 km $s^{-1}$
when we decrease $T_{eff}$ of 100 K.  
The two variations are basically symmetric and we chose as final error 
the absolute value of the largest one. 
Table \ref{errt} lists the differences between the 
new analysis and the original one for each abundance ratio. 
This method naturally includes both the errors due to the parameters and the 
covariance terms due to the interdependence between the parameters 
\citep[see also][for a complete discussion about the covariance terms]{mcw95}.

\section{Chemical abundance results}
\label{result}

\begin{figure}[h]
\plotone{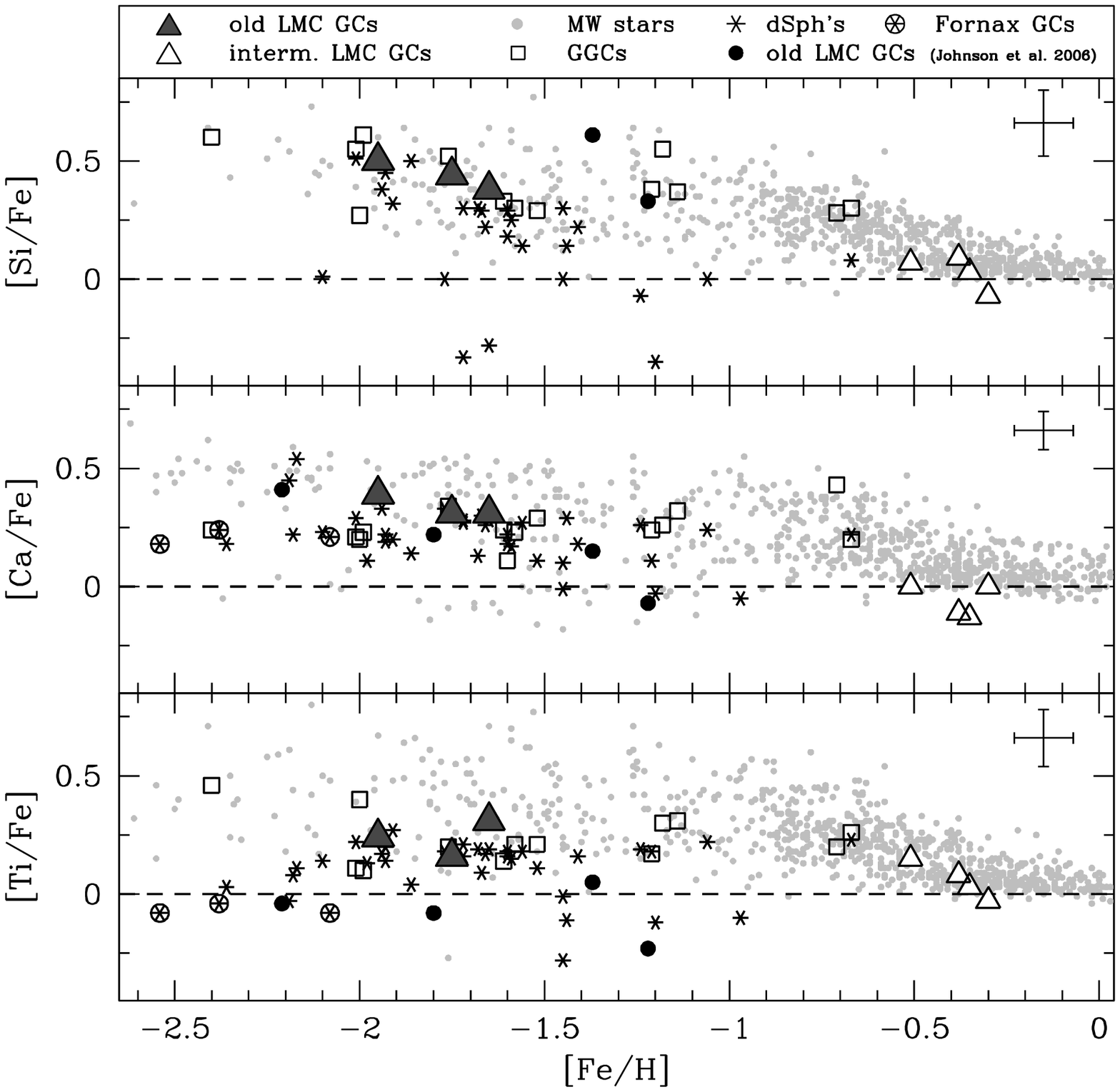}
\caption{Behavior of [Si/Fe], [Ca/Fe] and [Ti/Fe] abundance ratios as a 
function of [Fe/H]. The LMC clusters of this study are 
plotted as grey triangles and the results by \citet{m08} as 
white triangles.
Small grey points are Galactic stars. Empty squares 
are GGCs. Asteriks are dSphs field stars and Fornax GCs. Black points 
are the old LMC GCs by \citet{johnson06}.
All the references are in Table \ref{ref}.
Dashed lines mark the solar value.
The errorbar in the corner indicates 
the typical uncertainty associated to each abundance ratio and computed 
by summing in quadrature the internal error (reported in Tables~\ref{tabparam}-\ref{el3}) 
and the error from the adopted parameters (see Table~\ref{errt}). 
}
\label{alfa}
\end{figure}

\begin{figure}[h]
\plotone{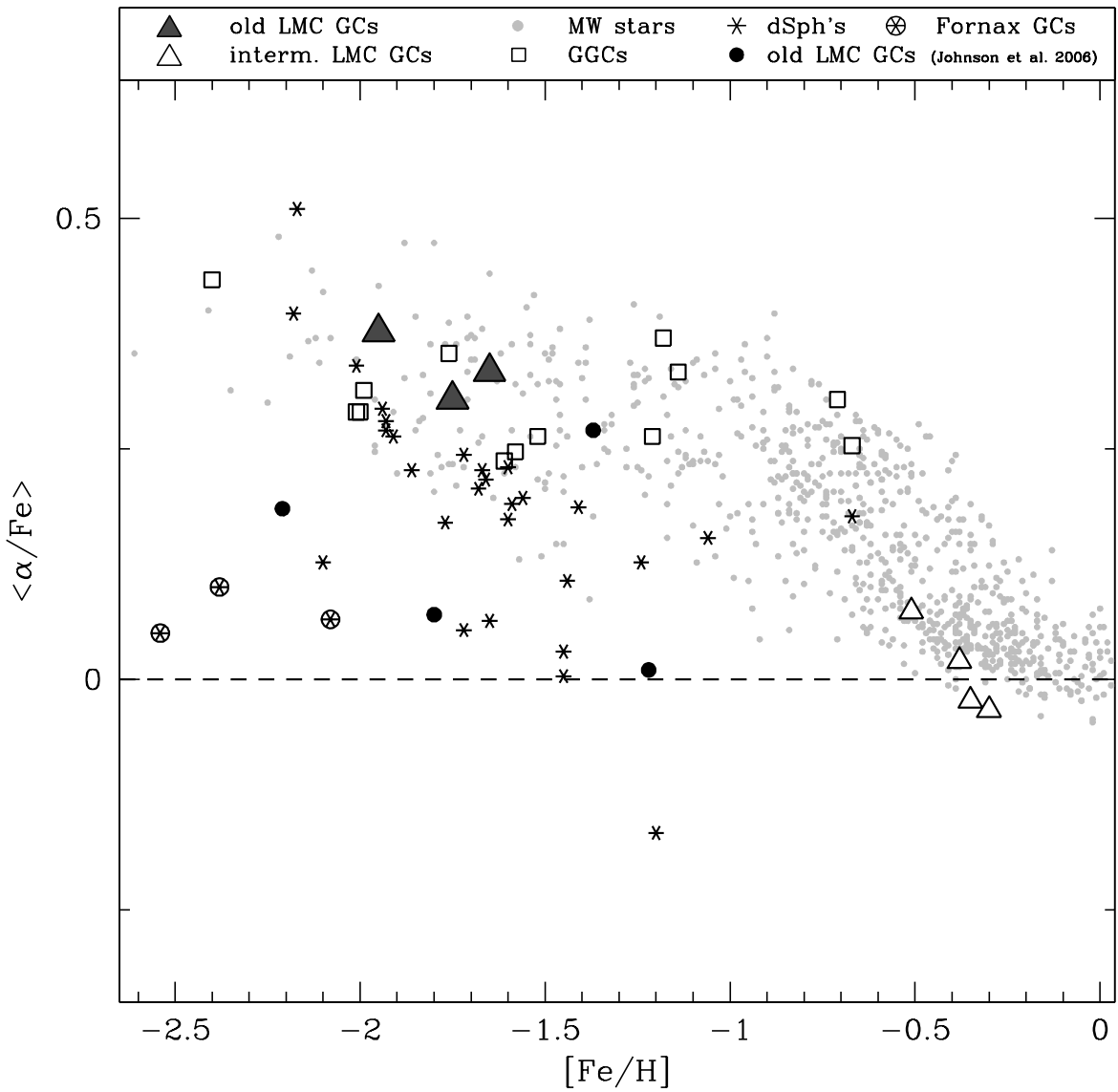}
\caption{Behavior of the average $<\alpha/Fe>$ ratio (defined as mean of 
[Si/Fe], [Ca/Fe] and [Ti/Fe]) as a function of [Fe/H].}
\label{alfa2}
\end{figure}

\begin{figure}[h]
\plotone{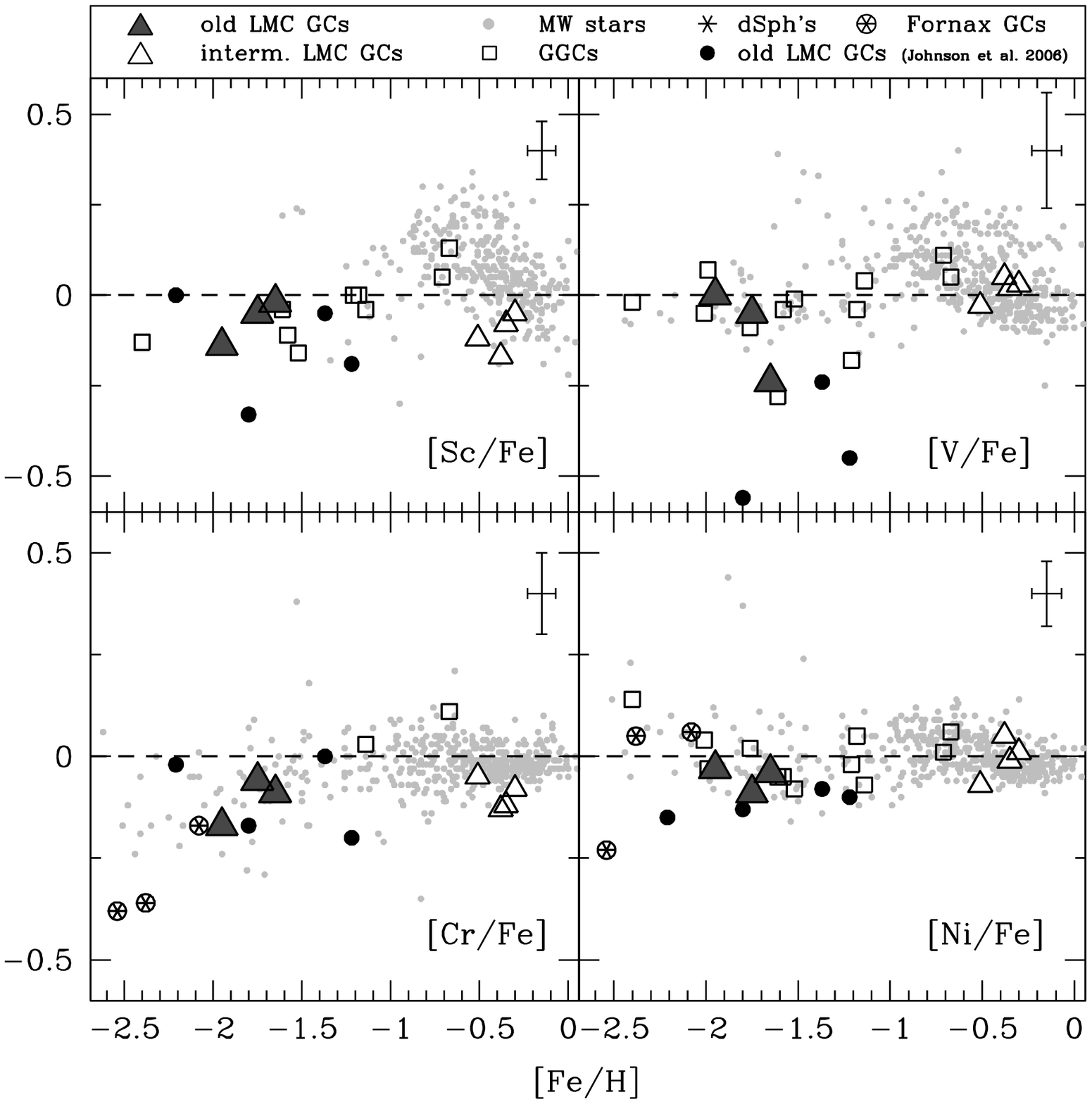}
\caption{Behavior of [Sc/Fe], [V/Fe], [Cr/Fe] and [Ni/Fe] 
as a function of [Fe/H]. }
\label{ironp}
\end{figure}

\begin{figure}[h]
\plotone{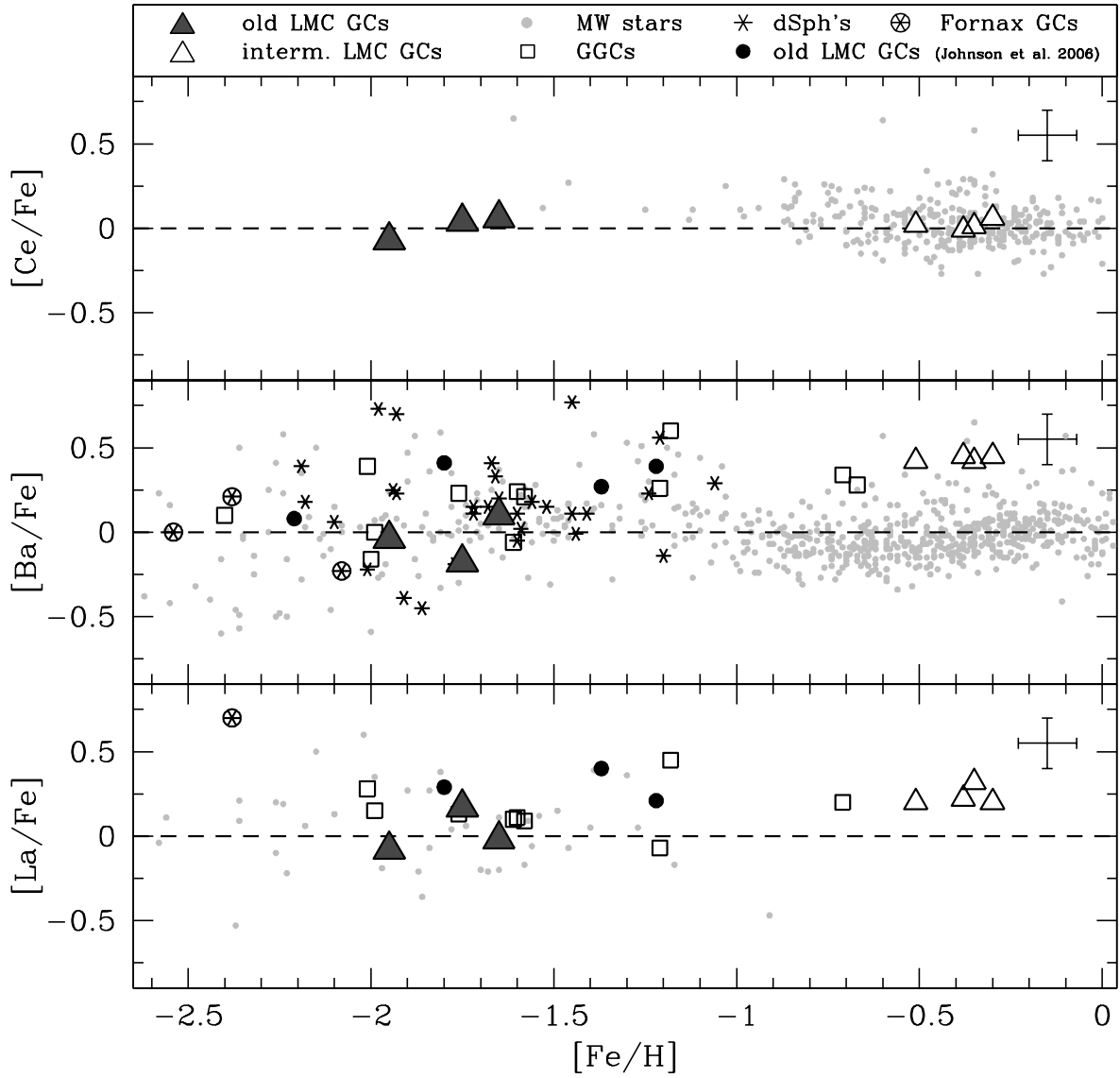}
\caption{Behavior of [Ce/Fe], [Ba/Fe] and [La/Fe]
as a function of [Fe/H]. }
\label{neu1}
\end{figure}

\begin{figure}[h]
\plotone{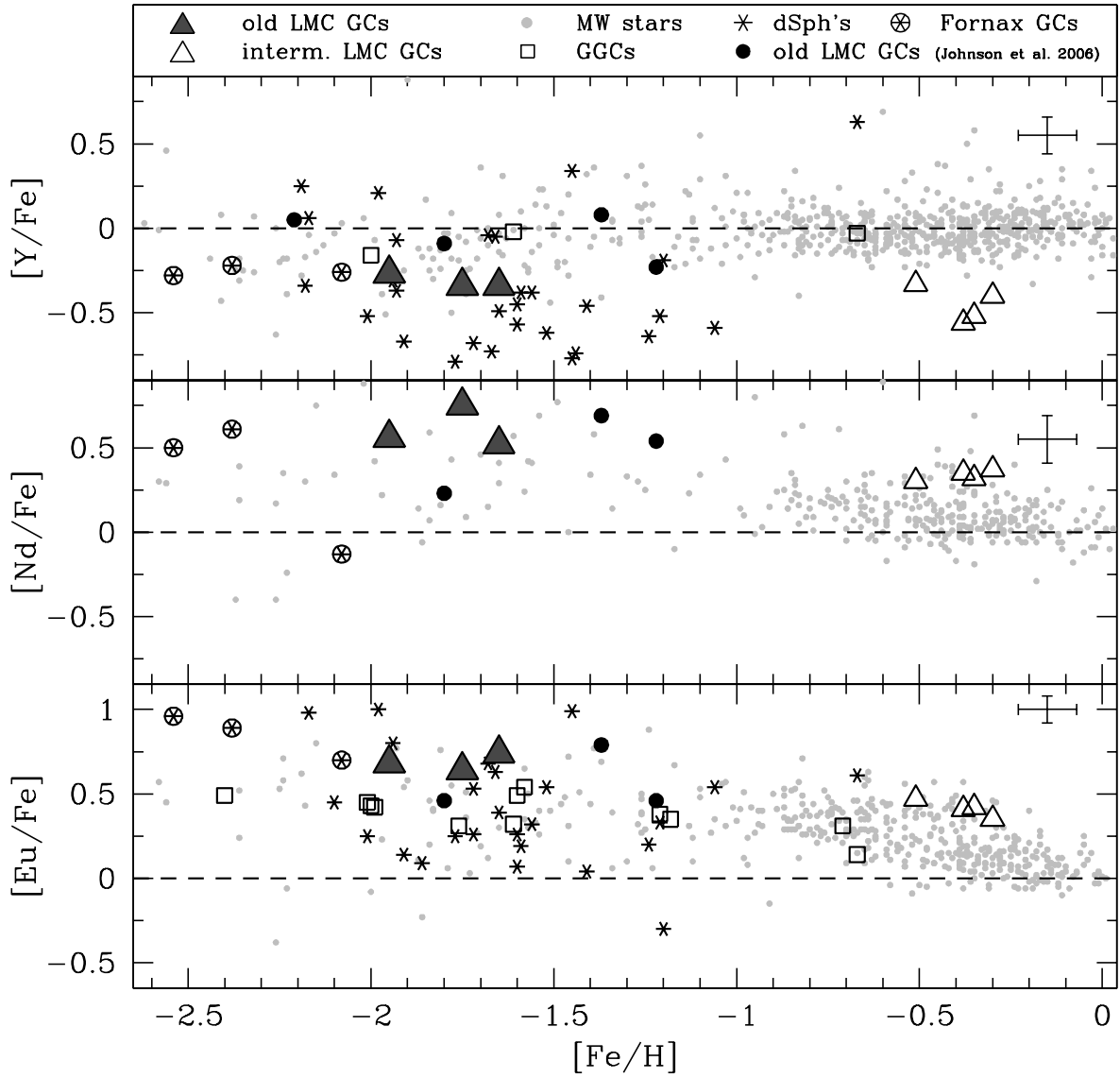}
\caption{Behavior of [Y/Fe], [Nd/Fe] and [Eu/Fe]
as a function of [Fe/H].}
\label{neu2}
\end{figure}

Tables \ref{el1} - \ref{el3} list the derived 
abundance ratios for all the 
studied stars. Table \ref{aver} summarizes the
cluster average abundance ratios, together with 
the dispersion around the mean. Figures \ref{alfa} - \ref{neu2} show
the plot of some abundance ratios
as a function of the iron content obtained 
in this work (as grey triangles) and in \citet{m08} 
(as white triangles). 
In these figures abundances obtained for
Galactic field stars (small grey circles), GGCs 
(squares), dSph's stars (asterisks) and for the sample of 
old LMC clusters by \citet{johnson06} (black points) 
are also plotted for comparison. 
All the reference sources 
are listed in Table  \ref{ref}.
For sake of homogeneity and in order to avoid possible 
systematic effects in the comparison, we perform a study of the 
oscillator strengths and adopted solar values of the comparison samples, 
aimed at bringing all abundances in a common system. 
Since our analysis is differential, we decide not to correct 
abundances derived with the same methodology 
\citep{edv,gratton03,reddy,reddy06}. 
All the other dataset have been re-scaled to our adopted 
oscillator strengths and solar values.
We compared oscillator strengths of  
lines in common with our list, finding, if any, negligible offsets 
(within $\pm$0.03 dex). Log~gf of the Ti~I lines adopted by 
\citet{ful}, \citet{shetrone01} 
and \citet{shetrone03} are 0.07 dex higher than ours, while 
log~gf of the Y~II lines by \citet{steph} results lower than ours by 
-0.09 dex.
The differences in the individual element solar values 
are small, typically less 
than 0.05 dex and generally the offsets of log~gf and solar values 
cancel out, with the only exception of the Ca abundances 
based on the solar value by \citet{ag89}, which turns out to be 
0.09 dex higher than ours.

The main abundance patterns are summarized as follows:

\begin{itemize}
\item {\bf Fe, O, Na, Mg and Al}---
Results about Fe, O, Na, Mg and Al of the target stars 
have been presented and discussed in \citet{m09}.
We derived an iron content of [Fe/H]=~--1.75$\pm$0.01 dex 
($\sigma$=~0.02 dex), ~--1.65$\pm$0.02 dex 
($\sigma$=~0.04 dex) and ~--1.95$\pm$0.02 dex 
($\sigma$=~0.04 dex) for NGC~1786, NGC~2210 and NGC~2257, respectively.\\
At variance with the other elements, Mg and Al
exhibit large star-to-star variations in each cluster, while 
similar dishomogeneities  have been found in the O content 
of NGC~1786 and 2257, and in the Na content of NGC~1786.
Such scatters are not compatible with the observational 
errors and indicate the presence of intrinsic variations. 
The same Na-O and Mg-Al anticorrelations observed in the GGCs 
have been found in these LMC clusters 
\citep[see Fig.~2 of][]{m09}.
Similar patterns have been already detected in the GGCs 
studied so far and they are generally interpreted 
in terms of a {\sl self-enrichment} process, where 
the ejecta of the primordial Asymptotic Giant Branch (AGB) 
stars (in which O and Mg have been destroyed producing 
large amount of Na and Al) are able to trigger the 
formation of a second stellar generation \citep{ventura01, ventura08}.
A complete discussion about the Na-O and Mg-Al anticorrelations in these 
3 LMC clusters is also presented in \citet{m09}.

\item {\bf $\alpha$-elements}---
Fig. \ref{alfa} shows the behavior of [Si/Fe], [Ca/Fe] 
and [Ti/Fe] as a function of [Fe/H] for the observed clusters and the 
comparison samples. The first 2 abundance ratios are enhanced, 
with [Si/Fe]$\sim$~+0.40 dex and [Ca/Fe]$\sim$~+0.30 dex, in
good agreement with the Halo and GGCs stars, while [Ti/Fe] 
is only moderately enhanced ($\sim$~+0.2 dex). 
Fig.~\ref{alfa2} shows the average of [Si/Fe], [Ca/Fe] and [Ti/Fe]
abundance ratios. We find $<\alpha/Fe>$ of 0.30$\pm$0.08, +0.33$\pm$0.02 
and +0.38$\pm$0.08 for NGC~1786, 2210 and 2257, respectively.
Such a level of $\alpha$-enhancement is consistent 
with that observed in the Galactic Halo ( both in field and cluster 
stars of similar metallicity), while dSphs 
display $<\alpha/Fe>$ ratios only $\sim$0.1-0.15 dex lower.
It is worth noticing that recent studies indicate that 
the $\alpha$-enhancement of the Sculptor stars well agrees with the Halo stars 
for lower metallicities \citep[see e.g.][]{tolstoy}, while the 
Fornax GCs show only a mild enhancement \citep{letarte}, see Fig.~\ref{alfa2}.
 
The only previous chemical analysis of $\alpha$-elements in 
old LMC GCs has been performed by \citet{johnson06}, analyzing 
4 GCs (namely, NGC~2005, 2019, 1898 and Hodge~11)
in the metallicity range [Fe/H]=~--2.2 / -1.2 dex 
(none of these objects is in common with our sample). 
At variance with us, they find solar or sub-solar [Ti/Fe] ratios and
moderately enhanced [Ca/Fe] ratios,
while their [Si/Fe] abundance ratios turn out to be enhanced  
in good agreement with our abundances.
However, we point out that the solar zero-point for their [Ca/Fe] 
(including both the 
solar reference for Ca and Fe) is +0.11 dex higher than ours. 
Taking into account this offset, their Ca abundances  are only
0.1 dex lower and still barely consistent
within the quoted uncertainties. Conversely, for Ti, the offset in 
the log~gf scale of -0.06 dex is not sufficient to erase the 
somewhat larger discrepancy ($\sim$0.2-0.3 dex) between the two abundance 
estimates.

\item {\bf Iron-peak elements}--- 
The abundance ratios for [Sc/Fe], [V/Fe], [Cr/Fe] and [Ni/Fe]  
are plotted in Fig.~\ref{ironp}. Such ratios turn out to be 
solar or (in a few case) moderately depleted, and consistent 
with the patterns observed in the Galactic Halo. The old LMC clusters 
analyzed by \citet{johnson06} exhibit similar abundance ratios, with 
the exception of [V/Fe] that appears to be depleted with respect to 
the solar value ([V/Fe]$<$--0.25 dex).
V is very sensitive to the adopted $T_{eff}$, as far as Ti, and 
we checked possible systematic offset between our $T_{eff}$ scale and that 
by \citet{johnson06}. Both scales are based 
on the excitational equilibrium, thus, the derived $T_{eff}$ are 
formally derived in a homogenous way. We checked possible offset in 
the adopted Fe log~gf, finding an average difference 
$log~gf_{J06}$-$log~gf_{this~work}$=~-0.004 ($\sigma$=~0.11). 
Moreover, there are no trends between the difference of the log~gf and 
$\chi$. We repeated our analysis for some stars by using the Fe log~gf 
by \citet{johnson06}, finding very similar $T_{eff}$ (within $\pm$50 K) 
with respect to our ones. Thus, we can consider that the two $T_{eff}$ scales 
are compatible each other. We cannot exclude that the different treatment 
of the hyperfine structure for the V~I lines between the two works be 
the origin of this discrepancy.
Unfortunately, we have no GCs in common with their sample and 
a complete comparison cannot be performed.

\item {\bf Neutron-capture elements}---
Elements heavier than the iron-peak (Z$>$31)
are built up  through rapid and slow neutron 
capture processes ({\sl r-} and {\sl s-}process, respectively).
Eu is considered a pure {\sl r}-process element, while
the first-peak {\sl s}-process element Y and 
the second-peak {\sl s}-process elements Ba, La, Ce and Nd 
(see Fig.~\ref{neu1} and \ref{neu2}), have an r contribution
less than $\sim$20-25\%  in the Sun.  
Nd is equally produced through {\sl s} and {\sl r}-process
\citep[see e.g.][]{arl99, burris}. 
Since the {\sl s}-process mainly occurs in AGB stars 
during the thermal pulse instability phase, 
s-process enriched gas should occur at later ($\sim$100-200 Myr) epochs.

In the measured old LMC clusters we find a general depletion ($\sim$~--0.30 dex) 
of [Y/Fe], still consistent (within the quoted uncertainties) with the 
lower envelope of the [Y/Fe] distribution of the Galactic stars, which 
show a solar-scaled pattern.
Also the metal-rich LMC clusters by \citet{m08} 
are characterized by such a depletion, with [Y/Fe] between --0.32 and 
--0.54 dex (see Fig.~\ref{neu2}).
Depleted [Y/Fe] ratios have been already
observed in dSphs field stars \citep{shetrone01, shetrone03} 
and in the Fornax GCs \citep{letarte}. 

The stars of NGC~2210 and NGC~2257 exhibit roughly solar [Ba/Fe] ratios 
(+0.10 and --0.04 dex, respectively), while 
in NGC~1786 this abundance ratio is depleted ([Ba/Fe]=~--0.18 dex). 
Also [La/Fe] and [Ce/Fe] show solar or slightly enhanced values, 
while [Nd/Fe] is always enhanced ($\sim$+0.50 dex).
The [Ba/Fe] ratio (as far as the abundances of other heavy {\sl s-}process 
elements) appears to be indistinguishable from the metal-poor stars in 
our Galaxy.

Fig.~\ref{neu2} (lower panel) shows the behavior of 
[Eu/Fe] as a function of the [Fe/H]. The 3 old LMC clusters exhibit 
enhanced ($\sim$~+0.7 dex) [Eu/Fe] ratios. These values are consistent 
with the more Eu-rich field stars in the Galactic Halo 
(that display a relevant star-to-star dispersion probably due to 
an inhomogeneous mixing), while 
the GGCs are concentrated around [Eu/Fe]$\sim$+0.40 dex
\citep{james}. The only other 
estimates of the [Eu/Fe] abundance ratio in LMC clusters have been provided 
by \citet{johnson06} who find enhanced values between $\sim$~+0.5 and +1.3 dex, 
fully consistent with our finding.

\end{itemize} 

\section{Discussion}

The $\alpha$-elements are produced mainly in the 
massive stars (and ejected via type II Supernovae 
(SNe) explosions) 
during both hydrostatic and explosive nucleosynthesis.  
As showed in Fig. \ref{alfa} and \ref{alfa2}, the LMC clusters 
of our sample display a behavior of [$\alpha$/Fe] 
as a function of [Fe/H] similar to the one observed in 
the Milky Way stars.
The enhanced [$\alpha$/Fe] ratios in the old LMC clusters suggest
that the gas from which these objects have been formed has been 
enriched by type II SNe ejecta on a relative short time-scale.
Such an observed pattern in the metal-poor regime agrees with 
the $\alpha$-enhancement of the Halo and GGCs stars, 
pointing out that the chemical contribution played by massive stars 
(concerning the nucleosynthesis of the $\alpha$-elements) in the
early epochs of the LMC and Milky Way has been similar.

\begin{figure}[h]
\plotone{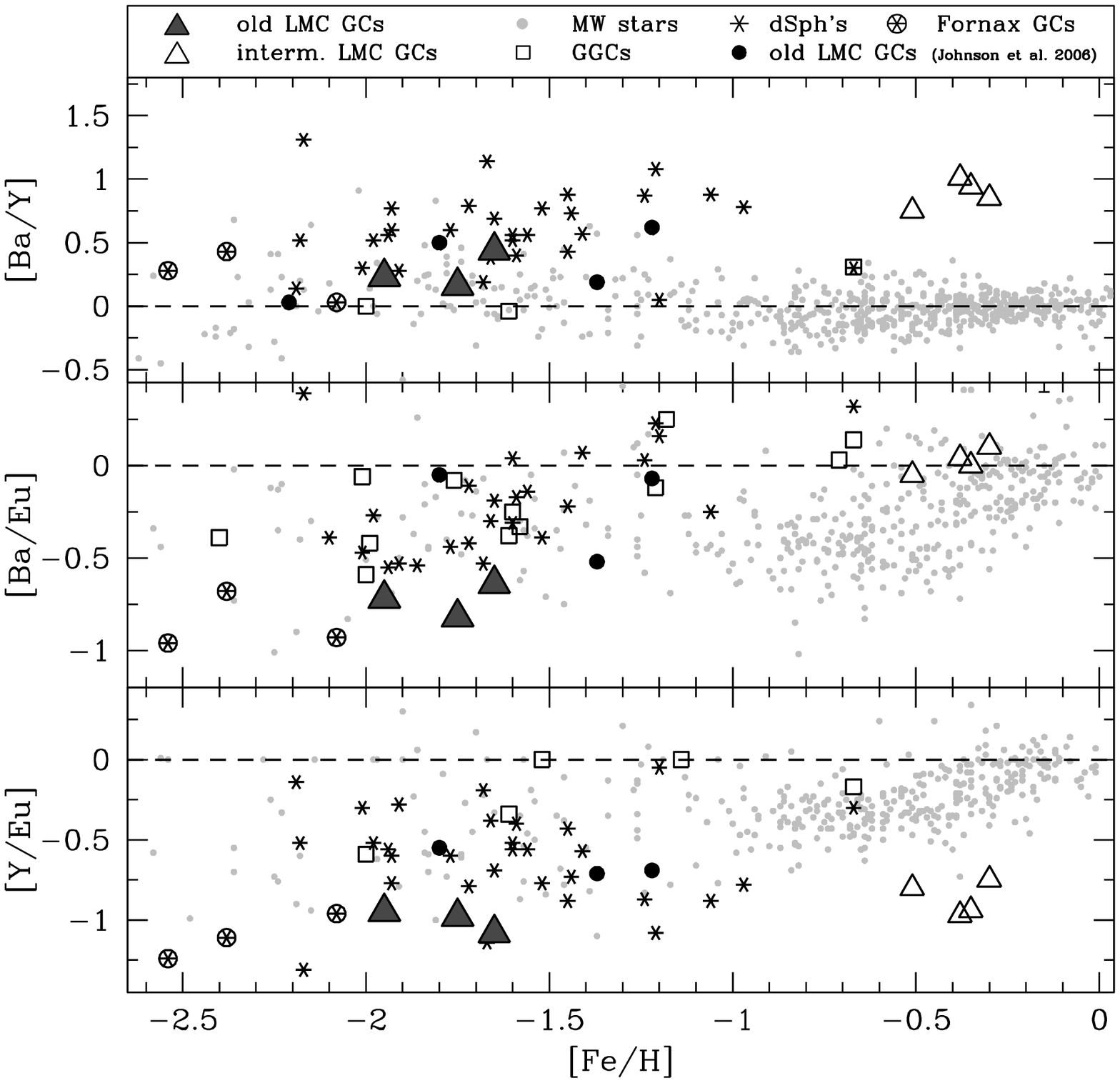}
\caption{Behavior of [Ba/Y], [Ba/Eu] 
and [Y/Eu] (lower panel) as a function of [Fe/H].}
\label{end1}
\end{figure}

[Ba/Y] is a convenient abundance ratio to estimate the relative contribution
between heavy and light {\sl s-}elements, [Ba/Eu] the relative contribution 
between heavy {\sl s} and {\sl r-}elements and [Y/Fe] the contribution 
between light {\sl s} and {\sl r-}elements. 
As shown in Fig.~\ref{end1} (upper panel), 
[Ba/Y] is solar or moderate enhanced in old LMC as in the Milky Way, 
but lower than the dSphs.
At higher metallicities the ratio increases appreciably due to 
the combined increase of Ba and decrease of Y. 
Such an increase  of [Ba/Y] with iron content can be ascribed to the rise 
of the AGB contribution, with a significant metallicity dependence of the AGB yields 
\citep[as pointed out by][]{venn}.

In the old LMC clusters, both the 
[Ba/Eu] and [Y/Eu] are depleted with respect to the solar value, 
with [Ba/Eu]$\sim$--0.70 dex and [Y/Eu]$\sim$--1 dex.
Such a depletion is consistent 
with the theoretical prediction by \citet{burris} and \citet{arl99} 
in the case of pure {\sl r}-process. 
Moreover, [Y/Eu] remains constant at all metallicities, at variance with 
[Ba/Eu] ratio. It is worth noticing that the precise nucleosynthesis 
site for Y is still unclear. 
Despite of the fact that most of the s-process elements are produced mainly 
in the He burning shell of intermediate-mass AGB stars, the lighter s-process
elements, such as Y, are suspected to be synthesized also during the central 
He burning phase 
of massive stars \citep[see e.g. the theoretical models proposed by][]{prantzos90}. 
Our results suggest that in the early ages of the LMC the nucleosynthesis of the 
heavy elements has been dominated by the {\sl r}-process, both because 
this type of process seems to be very efficient in the LMC 
and because the AGB stars have had no time to 
evolve and leave their chemical signatures in the interstellar medium.
The contribution 
by the AGB stars arises at higher metallicity (and younger age) when 
the AGB ejecta are mixed and their contribution becomes dominant. 
This hypothesis has been suggested also by \citet{shetrone03} in order to 
explain the lower [Y/Fe] abundance ratios observed in dSph's, pointing out 
a different Y nucleosynthesis for the Galaxy and 
the dSph's, with a dominant contribution by type II SNe in the Galactic satellites.

\begin{figure}[h]
\plotone{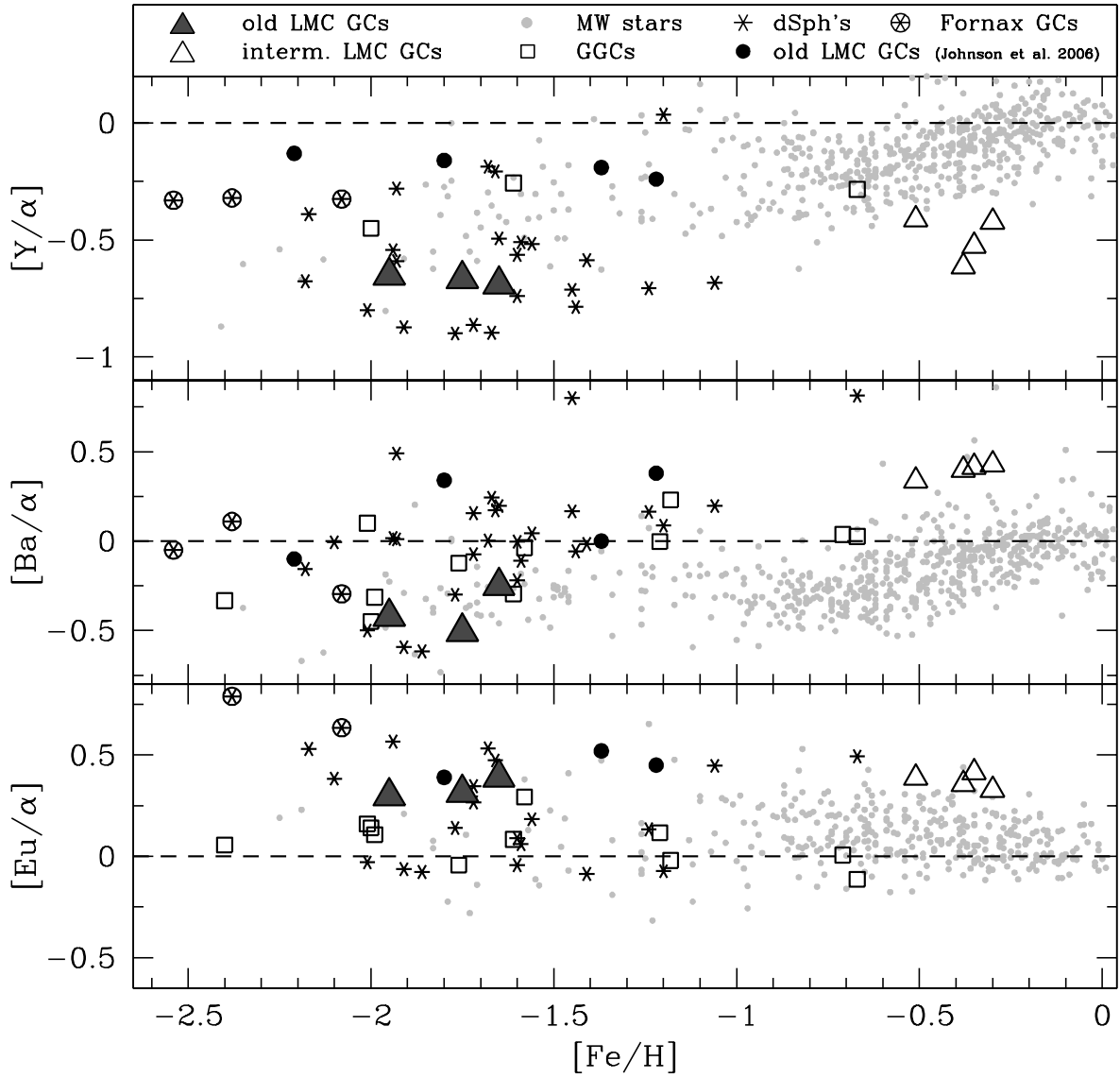}
\caption{Behavior of [Y/$\alpha$], [Ba/$\alpha$]
and [Eu/$\alpha$] as a function of [Fe/H].}
\label{end2}
\end{figure}

Fig.~\ref{end2} show the behaviour of [Y/$\alpha$], [Ba/$\alpha$] and 
[Eu/$\alpha$]. 
[Y/$\alpha$] and [Ba/$\alpha$] abundance ratios turns out to be depleted 
($<$--0.30 dex) at low metallicity, with a weak increase at higher metallicity 
for [Y/$\alpha$], while [Ba/$\alpha$] reaches $\sim$+0.50 dex.
This finding seems to confirm as Y is mainly produced by type II SNe, 
with a secondary contribution by low-metallicity AGB stars, 
at variance with Ba. In fact, in the low-metallicity AGB stars, the 
production of light s-process elements (as Y) is by-passed in favor 
to the heavy s-process elements (as Ba), because the number of seed nuclei 
(i.e. Fe) decrease decreasing the metallicity, while the neutron 
flux per nuclei seed increases.
In light of the spectroscopic evidences arising from our database of 
LMC GCs and from the previous studies about Galactic and dSphs stars, 
both irregular and spheroidal environments seem to share a similar 
contribution from AGB stars and type II SNe (concerning the neutron capture 
elements) with respect to our Galaxy. \\ 

Our LMC clusters sample shows a remarkably constant 
[Eu/$\alpha$] ratio of about $\sim$~+0.4 dex over the entire 
metallicity range, pointing toward  
a highly efficient {\sl r}-process mechanism
\footnote{
As a sanity check of our abundances in order to 
exclude systematic offset in the Eu abundances due to 
the adopted hyperfine treatment, we performed an analysis 
of [Eu/Fe] and [$\alpha$/Fe] ratios on Arcturus, by using 
an UVES spectrum taken from the UVES Paranal Observatory Project 
database \citep{bagnulo}. By adopting the atmospherical parameters 
by \citet{lecureur} and the same procedure described above,
we derived $<\alpha/Fe>$=~+0.23$\pm$0.09 dex, [Eu/Fe]=~+0.15$\pm$0.05 dex 
and [Eu/$\alpha$]=~--0.08 dex (according to the previous 
analysis by \citet{peterson93} and \citet{gopka}). 
For this reason, we exclude that the enhancement of 
[Eu/$\alpha$] in our stars can be due to an incorrect 
hyperfine treatment of the used Eu line.}.
First hints of such an enhanced [Eu/$\alpha$] pattern 
have been found in some supergiant stars in
the Magellanic Clouds \citep{hill95,hill99}, in Fornax GCs \citep{letarte}
and field stars (Bruno Letarte, Ph.D. Thesis) and in a bunch of Sgr stars 
\citep{boni00,mcw05}.

\section{Conclusion}
We have analyzed high-resolution spectra of 18 giants 
of 3 old LMC GCs, deriving abundance ratios 
for  13 elements, in addition to those already discussed 
in \citet{m09}  and sampling the different elemental groups, i.e. 
iron-peak, $\alpha$ and neutron-capture elements. 
The main results of our chemical analysis are summarized as follows:
\begin{itemize}
\item the three target clusters are metal-poor, with an iron content 
of [Fe/H]=~--1.75$\pm$0.01 dex 
($\sigma$=~0.02 dex), ~--1.65$\pm$0.02 dex 
($\sigma$=~0.04 dex) and ~--1.95$\pm$0.02 dex 
($\sigma$=~0.04 dex) for NGC~1786, NGC~2210 and NGC~2257, respectively 
 \citep[see][]{m09};
\item all the three clusters show the same level of enhancement 
of the $<\alpha/Fe>$ ratio ($\sim$~+0.30 dex), consistent 
with a gas enriched by type II SNe, while metal-rich, younger 
LMC clusters exhibit solar-scaled $<\alpha/Fe>$ ratio, due 
to the contribution of type Ia SNe at later epochs;
\item the iron-peak elements (Sc, V, Cr, Ni)  follow a solar pattern 
(or slightly sub-solar, in some cases), 
according with the observed trend in our Galaxy and consistent 
with the canonical nucleosynthesis scenario;
\item the studied clusters show a relevant ($\sim$--0.30 dex) depletion 
of [Y/Fe], while the other s-process elements (with the exception of Nd) 
display abundance ratios consistent with the Galactic distributions. 
[Ba/Fe] and [Ba/Y] in the old LMC GCs are lower than 
the values measured in the metal-rich, intermediate-age LMC GCs, because 
in the former the AGB stars had no time to evolve and enrich 
the interstellar medium;
\item ~[Eu/Fe] is enhanced ($\sim$+0.70 dex) in all the clusters.
This seems to suggest that the r-process elements production 
is very efficient in the LMC, being also the main channel of 
nucleosynthesis for the other neutron-capture elements.
\end{itemize}

In summary, the old, metal-poor stellar population of the LMC clusters 
closely resembles the GGCs in many chemical abundance patterns 
like the iron-peak, the $\alpha$ and heavy {\sl s-}process elements, and 
concerning the presence of chemical anomalies for Na, O, Mg and Al. 
When compared with dSphs the LMC old stellar population shows remarkably 
different abundance patterns for [$\alpha$/Fe] and neutron-capture elements.

\acknowledgements  
We warmly thank the anonymous referee for his/her useful comments.
This research was supported by the Ministero dell'Istruzione,
dell'Universit\'a e della Ricerca.

\newpage 

\begin{deluxetable}{lccccccc} 
\tablecolumns{8} 
\tablewidth{0pc}  
\tablecaption{Information about the target stars. S/N have been computed 
at 6000 $\mathring{A}$ for the UVES spectra and at 5720 and 6260 $\mathring{A}$ 
for the GIRAFFE HR~11 and 13 spectra respectively. RA and Dec are onto 2MASS 
astrometric system. Last column reports the adopted instumental configuration 
(U for UVES and G for GIRAFFE spectra).} 
\tablehead{ 
\colhead{Star ID}  & \colhead{$S/N$} & \colhead{$V_{helio}$}
 &  \colhead{$K_0$} & \colhead{$(J-K)_0$} & \colhead{RA(J2000)} & \colhead{Dec(J2000)} & \colhead{spectrum}\\
 &   & \colhead{(km/s)}& & & & &  } 
\startdata 
NGC~1786-978  &      ---~/~70~/~110	   & 260.5   &   13.55   & 0.78      &  74.7878641    & -67.7285246  &   G\\
NGC~1786-1248 &       45~/~---~/---	   & 255.4   &   13.50   & 0.77      &  74.7688292    & -67.7408723  & U \\
NGC~1786-1321 &       50~/~---~/---	   & 273.5   &   13.11   & 0.78      &  74.7638489    & -67.7546146  & U \\
NGC~1786-1436 &      ---~/~60~/~90	   & 267.1   &   13.71   & 0.72      &  74.7555606    & -67.7353347  &   G\\
NGC~1786-1501 &       40~/~---~/---	   & 265.9   &   12.92   & 0.93      &  74.7493142    & -67.7514295  & U  \\
NGC~1786-2310 &       50~/~---~/---	   & 262.2   &   12.83   & 0.82      &  74.7588569    & -67.7432595  & U \\
NGC~1786-2418 &      ---~/~70~/~100	   & 265.5   &   13.09   & 0.82  &  74.8215213    & -67.7387519  &   G\\
\hline
NGC~2210-122  &       40~/~---~/---	   & 337.7 &	13.22  &  0.75    & 92.9389070    & -69.1122894    &  U  \\
NGC~2210-309  &       40~/~---~/---	   & 338.4 &	13.29  &  0.75    & 92.9025764    & -69.1129818    &  U  \\
NGC~2210-431  &       50~/~---~/---	   & 340.0 &	13.04  &  0.77    & 92.8887909    & -69.1137252    &  U \\
NGC~2210-764  &       40~/~---~/---	   & 335.7 &	12.93  &  0.74    & 92.8575073    & -69.1267703    &  U  \\
NGC~2210-1181 &       50~/~---~/---	   & 335.6 &	12.81  &  0.77    & 92.8756190    & -69.1137519    &  U  \\
\hline
NGC~2257-136  & 	 40~/~---~/---     & 298.1	&   13.65  &  0.77    &  97.5823810    & -64.3262965  & U \\
NGC~2257-189  & 	--- ~/~70~/~90     & 299.6	&   13.54  &  0.77    &  97.5741597    & -64.3299382  & G	   \\
NGC~2257-295  & 	 35~/~---~/---     & 301.4	&   14.40  &  0.74    &  97.5615868    & -64.3159959  & U \\
NGC~2257-586  & 	--- ~/~50~/~60     & 300.6	&   14.36  &  0.70    &  97.5327178    & -64.3129344  & G	   \\
NGC~2257-842  & 	 45~/~---~/---     & 297.4	&   13.77  &  0.76    &  97.5591210    & -64.3394905  & U \\
NGC~2257-993  & 	--- ~/~70~/~90     & 298.9	&   13.49  &  0.81    &  97.4855884    & -64.3174261  & G	   \\
\enddata			
\label{info}							     
\end{deluxetable}


\begin{deluxetable}{lccccccc} 
\tablecolumns{8} 
\tablewidth{0pc}  
\tablecaption{ Atmospherical parameters and derived [Fe/H] ratio 
(with the number of used lines and the associated internal error defined as 
$\sigma/\sqrt{N_{lines}}$) for 
all the observed stars. Solar value for Fe is 7.54 dex \citep{gratton03}. Photometric 
temperatures (column 3) have been reported in comparison with the spectroscopic ones 
(column 2).}
\tablehead{ 
\colhead{Star ID}  & \colhead{$T_{eff}^{spec}$}& \colhead{$T_{eff}^{phot}$}& \colhead{log~g}
 &  \colhead{[A/H]} & \colhead{$v_t$} & \colhead{n} & \colhead{[Fe/H]} \\
& \colhead{(K)}& \colhead{(K)} & \colhead{(dex)}& &\colhead{(km/s)} & & \colhead{(dex)} }
\startdata 
NGC~1786-978  &  4250	&   4260   &   0.57 &  -1.75	  & 1.40     &    14   &   -1.73~$\pm$~0.02   \\
NGC~1786-1248 &  4280	&   4285   &   0.75 &  -1.75	 & 1.70      &    60   &   -1.74~$\pm$~0.02   \\
NGC~1786-1321 &  4250	&   4260   &   0.65 &  -1.75	 & 1.80      &    54   &   -1.73~$\pm$~0.01   \\
NGC~1786-1436 &  4420	&   4412   &   0.76 &  -1.75	 & 1.70      &    15   &   -1.76~$\pm$~0.02   \\
NGC~1786-1501 &  4100	&   3936   &   0.55 &  -1.80	 & 1.80      &    57   &   -1.79~$\pm$~0.01   \\
NGC~1786-2310 &  4100	&   4167   &   0.47 &  -1.75	 & 1.90      &    47   &   -1.72~$\pm$~0.01   \\
NGC~1786-2418 &  4160	&   4167   &   0.47 &  -1.80	 & 1.50      &    16   &   -1.75~$\pm$~0.02   \\
\hline
NGC~2210-122  &  4300	&   4334   &   0.60 &  -1.65	 & 1.70      &    31   &   -1.66~$\pm$~0.02   \\
NGC~2210-309  &  4250	&   4334   &   0.55 &  -1.70	 & 1.80      &    35   &   -1.69~$\pm$~0.03   \\
NGC~2210-431  &  4200	&   4285   &   0.70 &  -1.65	 & 1.80      &    46   &   -1.67~$\pm$~0.02   \\
NGC~2210-764  &  4270	&   4360   &   0.60 &  -1.60	 & 1.90      &    42   &   -1.58~$\pm$~0.02   \\
NGC~2210-1181 &  4200	&   4285   &   0.60 &  -1.60	 & 1.80      &    46   &   -1.64~$\pm$~0.02   \\
\hline
NGC~2257-136 &   4290	&   4285   &   0.65 &  -1.90	 & 1.95      &    38   &   -1.94~$\pm$~0.02   \\
NGC~2257-189 &   4290	&   4285   &   0.61 &  -1.90	 & 1.60      &    17   &   -1.92~$\pm$~0.02   \\
NGC~2257-295 &   4360	&   4360   &   0.96 &  -2.00	 & 1.50      &    40   &   -1.95~$\pm$~0.03   \\
NGC~2257-586 &   4480	&   4466   &   0.82 &  -2.00	 & 1.50      &    13   &   -1.92~$\pm$~0.03   \\
NGC~2257-842 &   4320	&   4309   &   0.95 &  -1.90	 & 1.50      &    39   &   -1.96~$\pm$~0.02   \\
NGC~2257-993 &   4200	&   4190   &   0.52 &  -2.00	 & 1.50      &    17   &   -2.02~$\pm$~0.03   \\
\hline
\enddata
\label{tabparam}
\end{deluxetable}


\setlength{\topmargin}{3pt}
\setlength{\textheight}{24cm}

\begin{landscape}
\begin{deluxetable}{lcccccccccccc} 
\tablecolumns{13} 
\tiny
\tablewidth{0pc}  
\tablecaption{[O/Fe], [Na/Fe], [Mg/Fe], [Al/Fe], [Si/Fe] and [Ca/Fe] abundance ratios 
for each observed stars with the number of used lines and the corresponding internal error.} 
\tablehead{ 
\colhead{Star ID}  & \colhead{n}& \colhead{[O/Fe]}
 &  \colhead{n} & \colhead{[Na/Fe]} & \colhead{n} & \colhead{[Mg/Fe]} & \colhead{n} & \colhead{[Al/Fe]} 
 & \colhead{n} & \colhead{[Si/Fe]} & \colhead{n} & \colhead{[Ca/Fe]} 
\\ 
\colhead{SUN}&  &\colhead{8.79}& &\colhead{6.21}& &\colhead{7.43}& &\colhead{6.23} 
& &\colhead{7.53} & & \colhead{6.27}}
\startdata 
NGC~1786-978     & 1 &  -0.15~$\pm$~0.12   & 3  &   0.47~$\pm$~0.03  & 1  &	 0.25~$\pm$~0.06 & ---  &   ---		     &  1  &  0.36$\pm$0.06  &    6    &	0.22$\pm$0.08	 \\
NGC~1786-1248    & 2 &   0.26~$\pm$~0.08   & 2  &   0.16~$\pm$~0.08  & 2  &	 0.51~$\pm$~0.08 & ---  &   $<$0.27	     &  3  &  0.24$\pm$0.07  &   14    &	0.32$\pm$0.02	 \\
NGC~1786-1321    & 2 &   0.31~$\pm$~0.07   & 2  &  -0.18~$\pm$~0.07  & 2  &	 0.41~$\pm$~0.07 & ---  &   $<$0.11	     &  3  &  0.49$\pm$0.06  &   17    &	0.23$\pm$0.03	   \\
NGC~1786-1436    & 1 &   0.18~$\pm$~0.09   & 1  &  -0.01~$\pm$~0.09  & 1  &	 0.40~$\pm$~0.09 & ---  &    ---	     &  1  &  0.57$\pm$0.09  &    5    &	0.37$\pm$0.07	 \\
NGC~1786-1501    & 2 &   0.30~$\pm$~0.08   & 4  &   0.60~$\pm$~0.06  & 1  &	 0.49~$\pm$~0.12 & 2    &   0.79~$\pm$~0.08  &  1  &  0.41$\pm$0.12  &   16    &	0.23$\pm$0.03	\\
NGC~1786-2310    & --- &   $<$-0.60        & 3  &   0.66~$\pm$~0.05  & 1  &	-0.21~$\pm$~0.08 & 2    &   1.02~$\pm$~0.06  &  4  &  0.51$\pm$0.04  &   14    &	0.40$\pm$0.03	  \\
NGC~1786-2418    & --- &  $<$-0.40         & 4  &   0.77~$\pm$~0.03  & 1  &	-0.31~$\pm$~0.07 & ---  &   ---		     &  1  &  0.52$\pm$0.07  &    5    &	0.39$\pm$0.04	 \\
\hline
NGC~2210-122     & 2 &   0.31~$\pm$~0.08   & 1  &  -0.08~$\pm$~0.11  & 1  &	 0.39~$\pm$~0.11 & ---  &   $<$0.54	     &  1  &   0.22$\pm$0.11 &   16    &	0.33$\pm$0.06	    \\
NGC~2210-309     & 1 &   0.10~$\pm$~0.14   & 4  &   0.69~$\pm$~0.10  & 1  &	 0.20~$\pm$~0.14 & 1    &   0.80~$\pm$~0.14  &  1  &   0.30$\pm$0.14 &   15    &	0.49$\pm$0.05	  \\
NGC~2210-431     & 2 &   0.12~$\pm$~0.11   & 3  &   0.64~$\pm$~0.07  & 1  &	 0.33~$\pm$~0.12 & 2    &   0.55~$\pm$~0.08  &  2  &   0.40$\pm$0.08 &   15    &	0.28$\pm$0.05	 \\
NGC~2210-764     & 2 &   0.25~$\pm$~0.10   & 2  &   0.32~$\pm$~0.10  & 1  &	 0.43~$\pm$~0.14 & ---  &   $<$0.30	     &  2  &   0.48$\pm$0.10 &   13    &	0.25$\pm$0.04	    \\
NGC~2210-1181    & 2 &   0.27~$\pm$~0.08   & 2  &  -0.03~$\pm$~0.08  & 2  &	 0.28~$\pm$~0.11 & ---  &   $<$0.20	     &  2  &   0.50$\pm$0.08 &   13    &	0.19$\pm$0.04	   \\
\hline
NGC~2257-136     & 1 &   0.22~$\pm$~0.11   & 2  &   0.20~$\pm$~0.11  & 1  &	 0.34~$\pm$~0.11 &  1   &   0.88~$\pm$~0.11  &  2	&   0.54$\pm$0.08      &   13    &    0.29$\pm$0.02    \\
NGC~2257-189     & --- &   $<$-0.20        & 2  &   0.49~$\pm$~0.07  & 1  &	 0.42~$\pm$~0.10 & ---  &    ---	     &  1       &	0.62$\pm$0.10  &    5	 &    0.37$\pm$0.04	\\
NGC~2257-295     & 1 &   0.24~$\pm$~0.18   & 3  &   0.58~$\pm$~0.10  & 1  &	 0.12~$\pm$~0.18 &  1   &   1.17~$\pm$~0.18  &  2	&   0.53$\pm$0.13      &   14    &    0.53$\pm$0.03    \\
NGC~2257-586     & --- &   $<$-0.20        & 2  &   0.22~$\pm$~0.08  & 1  &	 0.36~$\pm$~0.11 & ---  &    ---	     &  1       &	0.53$\pm$0.11  &    5	 &    0.31$\pm$0.05	\\
NGC~2257-842     & 1 &  -0.08~$\pm$~0.15   & 2  &   0.54~$\pm$~0.10  & 1  &	 0.52~$\pm$~0.15 &  --- &   $<$0.68	     &  2   	&   0.46$\pm$0.11      &   15    &    0.47$\pm$0.04 	\\
NGC~2257-993     & --- &   $<$-0.20        & 2  &   0.90~$\pm$~0.09  & 1  &	 0.24~$\pm$~0.13 & ---  &    ---	     &  1       &	0.34$\pm$0.13  &    5	 &    0.39$\pm$0.04    \\
\hline
\enddata
\label{el1}
\end{deluxetable} 
\end{landscape}

\begin{landscape}
\begin{deluxetable}{lcccccccccc} 
\tablecolumns{11} 
\tiny
\tablewidth{0pc}  
\tablecaption{[Ti/Fe], [Sc/Fe]~II, [V/Fe], [Cr/Fe] and [Ni/Fe]  abundance ratios 
for each observed stars with the number of used lines and the corresponding internal error.}
\tablehead{ 
\colhead{Star ID}  & \colhead{n}& \colhead{[Ti/Fe]}
 &  \colhead{n} & \colhead{[Sc/Fe]II} & \colhead{n} & \colhead{[V/Fe]} & \colhead{n} & \colhead{[Cr/Fe]} & \colhead{n} & \colhead{[Ni/Fe]} \\
\colhead{SUN}&  &\colhead{5.00}& &\colhead{3.13}& &\colhead{3.97}& &\colhead{5.67} & & \colhead{6.28} }
\startdata 
NGC~1786~978  &    3	&   0.11$\pm$0.03  &  3  & -0.04$\pm$0.03  &---    &	 ---		&---	 & ---  	   &  4  &   -0.04$\pm$0.03  \\
NGC~1786~1248 &   12	&   0.16$\pm$0.02  &  5  &  0.06$\pm$0.07  &   5   &	0.05$\pm$0.06	&   5	 & -0.03$\pm$0.05  & 10  &   -0.12$\pm$0.02  \\
NGC~1786~1321 &    9	&   0.13$\pm$0.02  &  5  & -0.17$\pm$0.07  &   7   &   -0.14$\pm$0.06	&   6	 & -0.11$\pm$0.04  & 11  &   -0.08$\pm$0.04    \\
NGC~1786~1436 &    4	&   0.40$\pm$0.05  &  4  & -0.14$\pm$0.05  &   1   &   -0.05$\pm$0.09	&---	 &---		   &  2  &   -0.09$\pm$0.06  \\
NGC~1786~1501 &   12	&   0.01$\pm$0.05  &  6  & -0.05$\pm$0.06  &   6   &   -0.18$\pm$0.08	&   5	 & -0.10$\pm$0.08  & 10  &   -0.11$\pm$0.05 \\
NGC~1786~2310 &   15	&   0.15$\pm$0.05  &  4  &  0.03$\pm$0.04  &   6   &	0.05$\pm$0.06	&   3	 &  0.00$\pm$0.05  & 12  &   -0.03$\pm$0.03   \\
NGC~1786~2418 &    2	&   0.13$\pm$0.05  &  4  & -0.03$\pm$0.04  &   1   &   -0.04$\pm$0.07	&---	 & ---  	   &  4  &   -0.14$\pm$0.04  \\
\hline
NGC~2210~122  &    6	&   0.38$\pm$0.08  &  5  & -0.05$\pm$0.07   &  2    &	-0.23$\pm$0.08  &   3	 &  -0.07$\pm$0.06    &   7 & -0.04$\pm$0.04	 \\
NGC~2210~309  &    9	&   0.35$\pm$0.06  &  4  &  0.12$\pm$0.07   &  5    &	-0.22$\pm$0.08  &   3	 &  -0.05$\pm$0.08    &   8 &  0.14$\pm$0.03   \\
NGC~2210~431  &    7	&   0.26$\pm$0.07  &  5  &  0.06$\pm$0.04   &  4    &	-0.09$\pm$0.07  &   6	 &  -0.04$\pm$0.08    &   7 & -0.15$\pm$0.05  \\
NGC~2210~764  &    7	&   0.26$\pm$0.09  &  6  & -0.19$\pm$0.06   &  5    &	-0.29$\pm$0.08  &   3	 &  -0.11$\pm$0.08    &  10 & -0.01$\pm$0.07	 \\
NGC~2210~1181 &    5	&   0.28$\pm$0.09  &  5  & -0.06$\pm$0.09   &  5    &	-0.35$\pm$0.03  &   3	 &  -0.16$\pm$0.06    &   7 & -0.14$\pm$0.08	\\
\hline
NGC~2257~136   &   8	& 0.24$\pm$0.05   &  6   &  -0.16$\pm$0.06  &  1    &	 -0.12$\pm$0.11 &   7	 &   -0.06$\pm$0.07   &  7  &  0.05$\pm$0.05	\\
NGC~2257~189   &   3	& 0.25$\pm$0.01   &  4   &  -0.19$\pm$0.05  &  ---  &---		&   ---  &   ---	      &   2 &  0.02$\pm$0.07				 \\
NGC~2257~295   &   4	& 0.33$\pm$0.08   &  4   &  -0.10$\pm$0.08  &  2    &	  0.13$\pm$0.06 &   4	 &   -0.28$\pm$0.08   &  5  & -0.11$\pm$0.04	\\
NGC~2257~586   &   ---  &   --- 	  &  3   &  -0.17$\pm$0.06  &  ---  &---		&---	 &   ---	      & 1   & -0.14$\pm$0.11				 \\
NGC~2257~842   &   9	& 0.24$\pm$0.06   &  6   &  -0.04$\pm$0.02  &  1    &	 -0.01$\pm$0.15 &   7	 &   -0.18$\pm$0.04   &  8  &  0.01$\pm$0.08	     \\
NGC~2257~993   &   3    & 0.16$\pm$0.05   &  4   &  -0.16$\pm$0.07  &  ---  &---                &---     &   ---              & 2   & -0.03$\pm$0.09  			    \\
\hline
\enddata
\label{el2}
\end{deluxetable} 
\end{landscape}

\setlength{\topmargin}{3pt}
\setlength{\textheight}{24cm}

\begin{landscape}
\begin{deluxetable}{lcccccccccccc} 
\tablecolumns{13} 
\tiny
\tablewidth{0pc}  
\tablecaption{[Y/Fe]~II, [Ba/Fe]~II, [La/Fe]~II, [Ce/Fe]~II, [Nd/Fe]~II and [Eu/Fe]~II  abundance ratios 
for each observed stars with the number of used lines and the corresponding internal error.}
\tablehead{ 
\colhead{Star ID}  & \colhead{n}& \colhead{[Y/Fe]II}
 &  \colhead{n} & \colhead{[Ba/Fe]II} & \colhead{n} & \colhead{[La/Fe]II}  &
  \colhead{n}& \colhead{[Ce/Fe]II}
 &  \colhead{n} & \colhead{[Nd/Fe]II} & \colhead{n} & \colhead{[Eu/Fe]II} 
\\ 
 \colhead{SUN}&  &\colhead{2.24}& &\colhead{2.13}& &\colhead{1.17}& &\colhead{1.58} & & \colhead{1.50} & & \colhead{0.51} 
 }
\startdata 
NGC~1786~978  & ---  & ---	       &  1   &   -0.21$\pm$0.12    &	 1   &  0.11$\pm$0.12	& ---  & ---		& ---	&   --- 	     &     ---    &  ---				  \\
NGC~1786~1248 &  3   & -0.36$\pm$0.09  &  3   &   -0.18$\pm$0.07    &	 1   &  0.01$\pm$0.12	& 1    & 0.08$\pm$0.12  &  3	&   0.65$\pm$0.07    &    1 &  0.60$\pm$0.12		  \\
NGC~1786~1321 &  2   & -0.48$\pm$0.08  &  3   &   -0.21$\pm$0.06    &	 1   &  0.32$\pm$0.10	& 1    & 0.11$\pm$0.10  &  3	&   0.85$\pm$0.06    &    1   &  0.78$\pm$0.10  	  \\
NGC~1786~1436 & ---  & ---	       &  1   &   -0.24$\pm$0.09    &  ---   &  ---		& ---  & ---		& ---	&  ---  	     &     ---    &  ---				  \\
NGC~1786~1501 &  1   & -0.20$\pm$0.12  &  3   &   -0.16$\pm$0.07    &	 1   &  0.24$\pm$0.12	& 1    &-0.13$\pm$0.12  &  3	&   0.87$\pm$0.07    &    1   &  0.69$\pm$0.12  	  \\
NGC~1786~2310 &  2   & -0.32$\pm$0.06  &  3   &   -0.06$\pm$0.05    &	 1   &  0.10$\pm$0.08	& 1    & 0.10$\pm$0.08  &  2	&   0.63$\pm$0.06    &    1 &  0.49$\pm$0.08		  \\
NGC~1786~2418 & ---  & ---	       &  1   &   -0.19$\pm$0.07    &	 1   &  0.26$\pm$0.07	& ---  & ---		& ---	& ---		     &     ---    &  ---				 \\
\hline
NGC~2210~122  &  2 &  -0.32$\pm$0.08 &  3 &  0.11$\pm$0.06     &    1	&  -0.12$\pm$0.11     &  1    &  0.10$\pm$0.11  &  3 &   0.65$\pm$0.06     &	     1   & 0.82$\pm$0.11	  \\
NGC~2210~309  &  1 &  -0.31$\pm$0.14 &  2 &  0.09$\pm$0.10     &   ---  &    ---	      &  ---  & ---		&  3 &   0.64$\pm$0.08     &	     1   & 0.70$\pm$0.14	  \\
NGC~2210~431  &  1 &  -0.40$\pm$0.12 &  3 &  0.07$\pm$0.07     &    1	&   0.08$\pm$0.12     &  1    &  0.07$\pm$0.12  &  3 &   0.56$\pm$0.07     &  1   & 0.77$\pm$0.12		\\
NGC~2210~764  &  2 &  -0.25$\pm$0.10 &  3 &  0.03$\pm$0.10     &    1	&   0.00$\pm$0.14     &  1    & -0.08$\pm$0.14  &  3 &   0.34$\pm$0.10     &	     1   & 0.75$\pm$0.14	  \\
NGC~2210~1181 &  2 &  -0.41$\pm$0.08 &  3 &  0.09$\pm$0.06     &    1	&  -0.06$\pm$0.11     &  1    &  0.15$\pm$0.11  &  3 &   0.43$\pm$0.06     &	     1   & 0.63$\pm$0.11	     \\
\hline
NGC~2257~136 &  2  &  -0.29$\pm$0.08  & 3  &  0.01$\pm$0.06 & 1  & $<$--0.10	&  1	 &  $<$0.00 & 3    & 0.71$\pm$0.06 & 1   &  0.75$\pm$0.11			     \\
NGC~2257~189 & --- &  ---	      & 1  & -0.06$\pm$0.10 & 1  & $<$--0.10	&  ---   &  ---     & ---  &  ---	   & --- &  --- 				    \\
NGC~2257~295 &  1  &  -0.28$\pm$0.18  & 3  & -0.07$\pm$0.10 & 1  & $<$0.00	&  1	 &  $<$0.10 & 4    & 0.48$\pm$0.09 & 1   &  0.59$\pm$0.18			     \\
NGC~2257~586 & --- &  ---	      & 1  & -0.11$\pm$0.11 & ---  & ---	&  ---   &  ---     & ---  &  ---	   & --- &  --- 				    \\
NGC~2257~842 &  2  &  -0.23$\pm$0.11  & 3  & -0.01$\pm$0.09 & 1  & $<$--0.10	&  1	 &  $<$0.10 & 3    & 0.50$\pm$0.09 & 1   &  0.70$\pm$0.15			      \\
NGC~2257~993 & --- &  ---             & 1  &  0.02$\pm$0.13 & 1  & $<$--0.10    &  ---   &  ---	& ---  &  ---	       & --- &  ---	     				 \\
\hline
\enddata
\label{el3}
\end{deluxetable} 
\end{landscape}

\begin{deluxetable}{lccc} 
\tablecolumns{4} 
\tablewidth{0pc}  
\tablecaption{Variation of each abundance ratio due to atmospherical 
parameters, obtained according to the method by \citet{cayrel04}. 
Second column reports the difference for each abundance ratio 
between the model with $T_{eff}$ increased by 100 K (and the re-optimization 
of the other parameters) and the original one. The third column reports 
the same differences but considering a model $T_{eff}$ decreased by 100 K. 
The last column lists the final average error.}
\tablehead{ 
\colhead{Ratio}  & \colhead{$(MOD)_{+100 K}$-MOD}& \colhead{$(MOD)_{-100 K}$-MOD} 
& \colhead{Average} \\
 &  \colhead{(dex)} & \colhead{(dex)}  & \colhead{(dex)} }
\startdata 
 $[O/Fe]$    	  &  +0.13   &	  --0.11    &	$\pm$0.12 \\
 $[Na/Fe]$   	  & --0.07   &     +0.06    &	$\pm$0.07   \\
 $[Mg/Fe]$   	  & --0.04   &     +0.05    &	$\pm$0.05   \\
 $[Al/Fe]$   	  & --0.05   &     +0.04    &	$\pm$0.05   \\
 $[Si/Fe]$   	  & --0.03   &     +0.10    &	$\pm$0.07   \\
 $[Ca/Fe]$   	  & --0.02   &     +0.01    &	$\pm$0.02   \\
 $[Sc/Fe]II$ 	  &  +0.06   &     +0.02    &	$\pm$0.04   \\
 $[Ti/Fe]$   	  &  +0.09   &    --0.10    &	$\pm$0.10    \\ 
 $[V/Fe]$    	  &  +0.11   &    --0.12    &	$\pm$0.12    \\ 
 $[Cr/Fe]$   	  &  +0.03   &    --0.06    &	$\pm$0.05    \\ 
 $[Fe/H]$    	  &  +0.08   &    --0.09    &	$\pm$0.09    \\ 
 $[Ni/Fe]$   	  &  +0.03   &    --0.02    &	$\pm$0.03    \\ 
 $[Y/Fe]II$  	  &  +0.02   &    --0.04    &	$\pm$0.04   \\ 
 $[Ba/Fe]II$ 	  &  +0.07   &    --0.09    &	$\pm$0.09    \\ 
 $[La/Fe]II$ 	  &  +0.15   &    --0.09    &	$\pm$0.15    \\  
 $[Ce/Fe]II$ 	  &  +0.09   &    --0.03    &	$\pm$0.06    \\  
 $[Nd/Fe]II$ 	  & --0.08   &     +0.11    &	$\pm$0.10    \\  
 $[Eu/Fe]II$ 	  &  +0.04   &    --0.03    &	$\pm$0.04    \\ 	    
\hline
\enddata
\label{errt}
\end{deluxetable}

\begin{deluxetable}{lcccccc} 
\tablecolumns{7} 
\tablewidth{0pc}  
\tablecaption{Average abundance ratios for the 3 old LMC clusters discussed in this study 
with the corresponding dispersion by the mean.} 
\tablehead{ 
\colhead{Ratio}  & \colhead{NGC~1786}& \colhead{}
 &  \colhead{NGC~2210} & \colhead{} & \colhead{NGC~2257} & \colhead{} \\
& \colhead{Mean} & \colhead{$\sigma$}&\colhead{Mean} & \colhead{$\sigma$}&\colhead{Mean} & \colhead{$\sigma$} }
\startdata 
$[O/Fe]$     &   $<$--0.04  &  0.36  &  0.23    &  0.07    &        $<$--0.06 & 0.18	 \\   
$[Na/Fe]$    &    0.22	 &  0.34  &  0.33    &  0.09    &	0.33    & 0.14	 \\ 
$[Mg/Fe]$    &    0.35	 &  0.36  &  0.31    &  0.36    &	0.46    & 0.29	 \\ 
$[Al/Fe]$    &    $<$0.55	 &  0.43  &     $<$0.48 &  0.23    &         $<$0.91  & 0.25 \\ 
$[Si/Fe]$    &   0.44 	 &  0.11  &  0.38    &  0.12    &	0.50    & 0.09	 \\ 
$[Ca/Fe]$    &   0.31 	 &  0.08  &  0.31    &  0.11    &	0.39    & 0.09 \\ 
$[Sc/Fe]II$  &  --0.05   &  0.08  & --0.02   &  0.12    &      --0.14    & 0.06 \\  
$[Ti/Fe]$    &    0.16	 &  0.12  &  0.31    &  0.05    &	0.24    & 0.06	 \\ 
$[V/Fe]$     &  --0.05	 &  0.09  & --0.24   &   0.10   &	 0.00   & 0.12 \\ 
$[Cr/Fe]$    &  --0.06   &  0.05  & --0.09   &   0.05   &      --0.17   & 0.11	 \\ 
$[Fe/H]$     &  --1.75 	 &  0.02  &  --1.65  &   0.04   &      --1.95    &	0.04 \\ 
$[Ni/Fe]$    &  --0.09 	 &  0.04  &  --0.04  &   0.12   &	--0.03    & 0.08	 \\ 
$[Y/Fe]II$   &  --0.34   &  0.11  &   --0.34 &   0.07   &	--0.27    & 0.03	 \\  
$[Ba/Fe]II$  &  -0.18    &  0.05  &    0.10  &    0.03  &	--0.04    & 0.05	 \\ 
$[La/Fe]II$  &   0.17    &  0.12  &  -0.02   &    0.08  &	$<$-0.08    &	0.04 \\ 
$[Ce/Fe]II$  &   0.04    &  0.11  &   0.06   &    0.10  &	$<$-0.07    &   0.06 \\ 
$[Nd/Fe]II$  &   0.75    &  0.12  &  0.52    &    0.14      &	0.56    & 0.13	 \\ 
$[Eu/Fe]II$  &   0.64    &  0.12  &  0.74    &    0.07      &	0.68    & 0.08	 \\ 
\hline
\enddata
\tablecomments{[Fe/H], [O/Fe], [Na/Fe], [Mg/Fe] and [Al/Fe] abundance ratios are from 
\citet{m09} and reported here for sake of completeness.}
\label{aver}
\end{deluxetable}

\begin{deluxetable}{lc} 
\tablecolumns{2} 
\tablewidth{0pc}  
\tablecaption{Literature sources for the comparison samples. } 
\tablehead{ 
\colhead{}  & \colhead{Reference} 
 }
\startdata
\hline
   &   Galactic GCs      \\
\hline
 47 Tuc  &  \citet{carretta04}, \citet{james}     \\
 NGC 2808  & \citet{carretta06}      \\
 NGC 6287  &  \citet{lee}   	\\
 NGC 6293  &  \citet{lee}   	\\
 NGC 6397  &  \citet{james}     \\
 NGC 6541  &  \citet{lee}   	\\
 NGC 6752  &  \citet{yong05}     \\
 M3  &   \citet{sneden04}    \\
 M4  &  \citet{ivans99}     \\
 M5  &   \citet{ivans01}    \\
 M10  &   \citet{kraft95}    \\
 M13  &   \citet{sneden04}     \\
 M15  &   \citet{sneden97}     \\
 M71  &   \citet{ramirez02}    \\
\hline
   &   Galactic Field Stars      \\
\hline
 Thin/Thick &  \citet{edv,koch}     \\   
 Halo &  \citet{burris}     \\   
 Halo/Thick &  \citet{ful}     \\  
 Halo/Thick & \citet{steph} \\
 Halo/Thick &  \citet{gratton03}     \\  
 Thin &  \citet{reddy}     \\  
 Thick &  \citet{reddy06}     \\  
\hline
 &    dSph      \\ 
\hline
 Draco & \citet{shetrone01}  \\ 
 Sextans & \citet{shetrone01}   \\ 
 Ursa Minor & \citet{shetrone01}  \\ 
 Sculptor  & \citet{shetrone03,geisler05}   \\
 Fornax  & \citet{shetrone03,letarte}   \\
 Carina  & \citet{shetrone03}   \\
 Leo~I  & \citet{shetrone03}   \\
\hline
\enddata
\label{ref}
\end{deluxetable}

\end{document}